\documentclass[aps, prc, twocolumn, showpacs, floatfix, amsmath,
amssymb, longbibliography, superscriptaddress]{revtex4-1}

\usepackage[utf8]{inputenc}
\usepackage[english]{babel}

\usepackage{graphicx}
\usepackage{dcolumn}
\usepackage{units}
\usepackage{hyperref}
\usepackage{multirow}
\usepackage{microtype}
\usepackage{braket}
\usepackage{mathtools}

\newcolumntype{d}[1]{D{.}{.}{#1}}

\graphicspath{{.}{./figs/}}

\newcommand{\nm}{\ensuremath{N_\mathrm{max}}}
\newcommand{\ntm}{\ensuremath{N_{\mathrm{max}}^{\mathrm{tot}}}}
\newcommand{\hw}{\ensuremath{\hbar\omega}}
\newcommand{\lir}{\ensuremath{L_\mathrm{IR}}}
\newcommand{\lirmax}{\ensuremath{L_\mathrm{IR,max}}}
\newcommand{\dlir}{\ensuremath{\Delta L_\mathrm{IR}}}
\newcommand{\luv}{\ensuremath{\Lambda_\mathrm{UV}}}
\newcommand{\Einf}{\ensuremath{E_\infty}}
\newcommand{\dEinf}{\ensuremath{\Delta E_\mathrm{IR}}}
\newcommand{\kinf}{\ensuremath{\kappa_\infty}}
\newcommand{\epsbar}{\ensuremath{\bar\epsilon}}
\newcommand{\epsNLO}{\ensuremath{\epsilon_\mathrm{NLO}}}
\newcommand{\Evar}{\ensuremath{E_\mathrm{var}}}
\newcommand{\dEinfmax}{\ensuremath{\Delta E_\mathrm{IR,max}}}
\def\nuc#1#2{\relax\ifmmode{}^{#1}{\protect\text{#2}}\else${}^{#1}$#2\fi}
\newcommand{\hnuc}[2]{\relax\ifmmode\prescript{#1}{\Lambda}{\protect\text{#2}}\else\(\prescript{#1}{\Lambda}{\text{#2}}\)\fi}
\newcommand{\clustermass}[1]{\ensuremath{M_{#1}}}
\newcommand{\LL}{\ensuremath{\mathcal{L}}}
\newcommand{\mN}{\ensuremath{m_N}}
\newcommand{\NN}{\ensuremath{N\!N}}
\newcommand{\NNN}{\ensuremath{N\!N\!N}}
\newcommand{\piN}{\ensuremath{\pi\!N}}
\newcommand{\YN}{\ensuremath{Y\!N}}
\newcommand{\YNN}{\ensuremath{Y\!N\!N}}
\newcommand{\vyn}{\ensuremath{V_{Y\!N}}}
\newcommand{\Amo}{A\text{-}1}
\newcommand{\Am}[1]{A\text{-}#1}
\newcommand{\Lam}{\ensuremath{\Lambda_{\NN}}}
\newcommand{\Tmax}{\ensuremath{T_{\rm Lab}^{\rm max}}}
\newcommand{\LamYN}{\ensuremath{\Lambda_{\YN}}}
\newcommand{\loyn}{Bonn--J\"{u}lich LO \YN{}}
\newcommand{\nnlosim}{NNLO\textsubscript{sim}}
\newcommand{\nnlosimstd}{\nnlosim(\Lam = \unit[500]{MeV}, \Tmax =
  \unit[290]{MeV})}
\newcommand{\DEsep}[2]{\ensuremath{\Delta\Esep[\hnuc{4}{H,He};{#1}]{#2}}}
\newcommand{\Esep}[2][]{\ensuremath{E_\mathrm{sep}^\mathrm{#2}\ifx\relax#1\relax\else\!\left({#1}\right)\fi}}
\newcommand{\ksep}[1]{\ensuremath{\kappa_\mathrm{sep}^\mathrm{#1}}}
\newcommand{\ksepth}{\ksep{th}}
\newcommand{\ksepexp}{\ksep{exp}}
\newcommand{\Esepth}{\Esep[]{th}}
\newcommand{\Esepexp}{\Esep[]{exp}}
\newcommand{\YNCSM}{Y-NCSM}

\newcommand{\newabbreviation}[3]{\newcounter{#1}\expandafter\newcommand\csname#1\endcsname[1][]{\ifthenelse{\equal{##1}{abreviate}}{#2}{\ifthenelse{\equal{##1}{fullname}}{#3}{\ifthenelse{\equal{##1}{explain}}{#3 (#2)\stepcounter{#1}}{\ifthenelse{\value{#1}=0}{#3##1 (#2##1)\stepcounter{#1}}{#2##1}}}}}}
\newabbreviation{HO}{\text{HO}}{harmonic oscillator}
\newabbreviation{NCSM}{\text{NCSM}}{no-core shell model}
\newabbreviation{IR}{\text{IR}}{infrared}
\newabbreviation{UV}{\text{UV}}{ultraviolet}
\newabbreviation{LEC}{\text{LEC}}{low-energy constant}
\newabbreviation{LO}{\text{LO}}{leading order}
\newabbreviation{NLO}{\text{NLO}}{next-to-leading order}
\newabbreviation{NNLO}{\text{NNLO}}{next-to-next-to-leading order}
\newabbreviation{CoM}{\text{CM}}{center-of-mass}
\newabbreviation{EFT}{EFT}{effective field theory}
\newabbreviation{chiEFT}{$\chi$EFT}{chiral effective field theory}
\newabbreviation{PDF}{\text{PDF}}{probability density function}
\newabbreviation{MLE}{\text{MLE}}{maximum likelihood estimation}
\newabbreviation{MCMC}{\text{MCMC}}{Markov Chain Monte Carlo}
\newabbreviation{GP}{\text{GP}}{Gaussian Process}
\newabbreviation{CSB}{\text{CSB}}{charge symmetry breaking}
\newabbreviation{SRG}{\text{SRG}}{similarity renormalization group}

{}
\DeclareMathOperator{\pr}{pr}
\newcommand{\given}{\,|\,}

\begin{document}

\title{Nuclear physics uncertainties in light hypernuclei}

\author{D.~Gazda}
\email{Corresponding author: gazda@ujf.cas.cz}
\affiliation{Nuclear Physics Institute of the Czech Academy of Sciences,
  25068 Řež,
  Czech Republic}
\affiliation{Department of Physics,
  Chalmers University of Technology,
  SE-412 96 G\"oteborg,
  Sweden}

\author{T.~Yadanar~Htun}
\affiliation{Department of Physics,
  Chalmers University of Technology,
  SE-412 96 G\"oteborg,
  Sweden}
\affiliation{
  School of Physics and Center of Excellence in High Energy Physics and Astrophysics,
  Suranaree University of Technology,
  Nakhon Ratchasima 30000,
  Thailand}
\affiliation{Department of Physics,
  University of Mandalay,
  05032 Mandalay,
  Myanmar}

\author{C.~Forss\'en}
\affiliation{Department of Physics,
  Chalmers University of Technology,
  SE-412 96 G\"oteborg,
  Sweden}

\begin{abstract}
  The energy levels of light hypernuclei are experimentally accessible
  observables that contain valuable information about the interaction
  between hyperons and nucleons. In this work we study strangeness
  $S = -1$ systems \hnuc{3,4}{H} and \hnuc{4,5}{He} using the
  \emph{ab~initio} no-core shell model (NCSM) with realistic
  interactions obtained from chiral effective field theory
  ($\chi$EFT). In particular, we quantify the finite precision of
  theoretical predictions that can be attributed to nuclear physics
  uncertainties. We study both the convergence of the solution of the
  many-body problem (method uncertainty) and the regulator-{} and
  calibration data-dependence of the nuclear $\chi$EFT Hamiltonian
  (model uncertainty). For the former, we implement infrared
  correction formulas and extrapolate finite-space NCSM results to
  infinite model space. We then use Bayesian parameter estimation to
  quantify the resulting method uncertainties. For the latter, we
  employ a family of 42 realistic Hamiltonians and measure the
  standard deviation of predictions while keeping the leading-order
  hyperon--nucleon interaction fixed. Following this procedure we find
  that model uncertainties of ground-state \(\Lambda\) separation
  energies amount to $\sim \unit[20(100)]{keV}$ in
  \hnuc{3}{H}(\hnuc{4}{H,He}) and $\sim\unit[400]{keV}$ in
  \hnuc{5}{He}. Method uncertainties are comparable in magnitude for
  the \hnuc{4}{H,He} \(1^+\) excited states and \hnuc{5}{He}, which
  are computed in limited model spaces, but otherwise much smaller.
  This knowledge of expected theoretical precision is crucial for the
  use of binding energies of light hypernuclei to infer the elusive
  hyperon--nucleon interaction.
\end{abstract}


\maketitle

\section{Introduction}
\label{sec:introduction}
One of the main goals of hypernuclear physics is to establish a
reliable link between the low-energy properties of hypernuclei and the
underlying nuclear and hyperon--nucleon (\YN) interactions. Unlike in
the nuclear sector, with its vast database of measured
nucleon--nucleon (\NN) scattering observables, the experimental data
on \YN{} scattering are unfortunately poorer both in quality and
quantity. Scattering experiments with hyperons are rather difficult to
perform due to their short life time. Consequently, bound states of
light hypernuclei play an essential and complementary role for our
understanding of \YN{} interactions. Over the past decades, the
structure of hypernuclei has been extensively studied worldwide---at
international facilities such as J-PARC, Jlab, CERN, and
FAIR---providing a great deal of precise information on binding
energies, as well as excitation spectra and even transition
strengths~\cite{Davis2005, Hashimoto2006, Gal2008, Gal:2012zz,
  Gibson:2010zz, Julia-Diaz:2013lla, Proceedings:2017nlf,
  Proceedings:2019rcl}.

These experimental efforts have been accompanied, and often driven, by
equally vigorous theory developments. In the past, several
phenomenological approaches have been employed to study hypernuclei,
such as the shell model for $p$-{} and $sd$-shell
hypernuclei~\cite{Gal1971, Gal1972, Gal1978, Millener2008,
  Millener2010, Millener2012}, various cluster~\cite{Motoba1983,
  Motoba1985, Hiyama2009, Hiyama2012} and mean-field
models~\cite{Glendenning1993, Vretenar1998, Vidana2001,
  Haidenbauer:2019thx}, as well as recent advanced Quantum Monte Carlo
calculations with simplified microscopic
interactions~\cite{Lonardoni2013, Lonardoni2014, Lonardoni2015}. Very
importantly, so-called ab~initio methods---capable of solving the
many-body Schr\"{o}dinger equation with controllable
approximations---have emerged very recently~\cite{Wirth2014,
  Wirth:2017bpw, Le:2019gjp, Le:2020zdu}. Such methods utilize
realistic hypernuclear Hamiltonians that are typically constrained to
describe \NN{} and \YN{} interactions in free space, and can include
three-body forces. Employing such interactions, the energy levels of
$A = 3, 4$ light hypernuclei have been calculated by solving Faddeev
and Yakubovsky equations already in~\cite{Miyagawa:1995sf, Nogga2002}.
The rapid advancements in theoretical many-body techniques, as well as
increased computing power, have recently paved the way to extend
ab~initio studies from $s$ shell up to the $p$-shell hypernuclei using
the \NCSM{} approach~\cite{Wirth2014, Wirth:2017bpw, Wirth:2017lso,
  Le:2019gjp}. This approach therefore provides an essential
cornerstone which allows us to asses the performance of available
microscopic hypernuclear Hamiltonians by confronting them with the
precise data on hypernuclear spectroscopy. Ab~initio calculations have
already proven to be a powerful tool, e.g., to reveal deficiencies in
available \YN{} interaction models \cite{Wirth2014}, as well as to
elucidate some of the long-standing question in hypernuclear physics,
such as the \CSB{} \cite{Gazda:2015qyt, Gazda:2016qva,
  Haidenbauer:2021wld} and the hyperon puzzle in dense nuclear matter
\cite{Wirth2016}. Furthermore, the use of ab initio methods promises
an opportunity to perform a rigorous quantification of theoretical
uncertainties. Showcasing such efforts is the main goal of the present
paper.

A prerequisite for ab~initio hypernuclear structure calculations is
the Hamiltonian, constructed from particular models of nuclear and
hypernuclear interactions. The current state of the art of the theory
of nuclear forces employs SU(2) \chiEFT{} formulated in terms of pions
and nucleons as the relevant degrees of freedom. It incorporates the
underlying symmetries and the pattern of spontaneous symmetry breaking
of QCD~\cite{PhysRevC.49.2932, RevModPhys.81.1773, MACHLEIDT20111}.
The fast pace of development in this field is reflected in the recent
emergence of plethora of different nuclear interaction
models~\cite{Carlsson:2015vda,Ekstrom:2015rta,LENPIC:2018ewt,Huther:2019ont,Jiang:2020the}.
Similarly, the development of \YN{} interactions has a rich history of
phenomenological models based on quark model~\cite{Fujiwara1996,
  Fujiwara1996a} and boson-exchange potentials~\cite{Rijken:2010zzb,
  Rijken:2006ep, Haidenbauer2005}. In recent years; however, \EFT{}
methods have been applied in the strangeness baryon--baryon sector as
well. Here, the pseudoscalar \(\pi\), \(K\) and \(\eta\) mesons;
together with the SU(3) octet baryons \(\Lambda\), \(\Sigma\) and
\(\Xi\) are the relevant degrees of freedom. Hyperon--nucleon forces
have been constructed employing SU(3) \chiEFT{} at
\LO{}~\cite{Polinder:2006zh} and \NLO{}~\cite{Haidenbauer2013,
  Haidenbauer:2019boi}, as well as using an alternative scheme
developed in~\cite{Baru:2019ndr, Ren:2019qow}. Alternatively, at very
low energies, even the pionic degrees of freedom can be integrated
out, resulting in the so-called pionless \EFT{}~\cite{BEDAQUE1998221,
  VANKOLCK1999273} which has been successfully applied to study light
nuclei~\cite{Hammer:2019poc} and even
hypernuclei~\cite{Contessi:2018qnz}. At the same time, lattice-QCD
calculations are expected to provide direct information on nuclear and
hypernuclear interactions in the near future. Especially in the
strangeness sector, where the experimental information is scarce or
does not exist, lattice QCD could provide valuable theoretical
information~\cite{Beane2012, Sasaki:2015ifa}.

The Hamiltonian itself is the main source of uncertainty in
calculations of light hypernuclei. The \YN{} interaction is rather
poorly constrained by the sparse \YN{} scattering database,
additionally suffering from large experimental uncertainties. As a
result, \YN{} interaction models differ already at the level of phase
shifts and this ambiguity leads to substantial uncertainties in
predictions of hypernuclear observables~\cite{Nogga2013, Wirth2014,
  Haidenbauer:2019boi}. On the other hand, this situation offers an
opportunity to utilize bound-state observables of light hypernuclei to
constrain the \YN{} interaction~\cite{Haidenbauer:2021wld}. In order
for such a program to be successful it is important to study all
relevant sources of uncertainty that enter when solving a many-body
problem with both nucleonic and hyperonic degrees of freedom. In
particular, the remnant freedom in the construction of realistic \NN{}
and \NNN{} interactions represents an additional source of model
uncertainty that enter in the prediction of hypernuclear properties.
Furthermore, the solution of the many-body problem using a truncated
\NCSM{} basis implies a method uncertainty that might become large for
increasing mass number.

The main purpose of this work is to quantify the theoretical precision
of relevant hypernuclear observables that can be attributed to nuclear
model and method uncertainties. More specifically, we study light
$\Lambda$ hypernuclei \hnuc{3,4}{H}, \hnuc{4,5}{He} using the \NCSM{}
with realistic Hamiltonians derived from \chiEFT{}. We quantify both
the theoretical model uncertainties due to the nuclear Hamiltonian and
the method errors from the solution of the many-body problem in
truncated bases. For the former we employ a family of 42 \nnlosim{}
realistic Hamiltonians obtained from \chiEFT{}~\cite{Carlsson:2015vda}
and study the sensitivity of the choice of interaction model on the
hypernuclear binding energies. For the latter we implement \IR{}
correction formulas to extrapolate the finite-space \NCSM{} results to
infinite model space, and we use Bayesian parameter estimation to
quantify the resulting method uncertainties.

The article is organized as follows: In Sec.~\ref{sec:method} we first
introduce the Jacobi-coordinate \NCSM{} (here denoted \YNCSM) for
hypernuclear systems with particles of unequal masses. We also
demonstrate the equivalence between the \YNCSM{} \HO{}-basis
truncation and \IR{}/\UV{} cutoff scales and we introduce the
realistic nuclear and \YN{} interactions from \chiEFT{} In
Sec.~\ref{sec:bayesIR}, we generalize the nuclear \IR{} correction
formulas~\cite{Forssen:2017wei} to hypernuclei and present a novel
Bayesian parameter inference method to extrapolate the \YNCSM{}
calculations to infinite model spaces. Results for ground-{} and
excited-state energies of \hnuc{3,4}{H}, \hnuc{4,5}{He} hypernuclei
are presented in Sec.~\ref{sec:results} and the consequences of our
findings are discussed in Sec.~\ref{sec:conclusion}.

\section{Method}
\label{sec:method}

\subsection{The hypernuclear no-core shell model}
\label{subsec:ncsm}
We employ the \textit{ab~initio} \NCSM{}~\cite{navratil2009} to solve
the many-body Schr\"{o}dinger equation. This nuclear technique was
extended recently to light hypernuclei~\cite{Wirth:2017bpw}. In
particular, we use the translationally invariant formulation of
\NCSM{} which involves a many-body \HO{} basis defined in relative
Jacobi coordinates~\cite{Navratil:1999pw}. For completeness, we
provide a short summary of the \YNCSM{} method in the following, but
refer to Ref.~\cite{Wirth:2017bpw} for details.

\YNCSM{} calculations start with the Hamiltonian for a system of
nonrelativistic nucleons and hyperons ($\Lambda$ and $\Sigma$)
interacting by \NN{}, three-nucleon (\NNN{}) and \YN{} interactions:
\begin{equation}
  \label{eq:ncsm:h}
  \begin{split}
    H &= \sum_{i=1}^A
        \frac{\vec{p}^{\,2}_{i}}{2m_i}
    + \sum_{i=1}^{A-1} V_{\YN,iA} + \Delta M \\
    &+ \sum_{1 \le i < j}^{A-1} V_{\NN,ij}
    + \sum_{1 \le i < j < k}^{A-1} V_{\NNN,ijk}.
  \end{split}
\end{equation}
The masses $m_i$, momenta $\vec{p}_i$, and indices \(i, j, k\)
correspond to the nucleonic degrees of freedom for \(i,j,k \le A-1\)
and for $i = A$ to hyperons. Since the \YN{} interaction model
employed in this work explicitly takes into account the
strong-interaction \(\Lambda N \leftrightarrow \Sigma N\) transitions,
the $\Lambda$-hypernuclear states are coupled with
$\Sigma$-hypernuclear states. To account for the mass difference of
these states, the mass term
\begin{equation}
  \label{eq:ncsm:deltam}
  \Delta M = \sum_{i\le A} m_i - M_0,
\end{equation}
is introduced in the Hamiltonian~\eqref{eq:ncsm:h}. Here, $M_0$ is the
reference mass of a hypernuclear system containing only nucleons and a
$\Lambda$ hyperon.

The Jacobi-coordinate formulation fully exploits the symmetries of the
Hamiltonian to decouple the \CoM{} motion and to construct an
angular-momentum-{} and isospin-coupled \HO{} basis states. In
\YNCSM{}, several different equivalent sets of Jacobi coordinates are
employed. The set
\begin{equation}
  \label{eq:ncsm:jsetxi}
  \begin{split}
    \vec{\xi}_0 &= \frac{1}{\sqrt{\clustermass{A}}}\sum_{i=1}^A \sqrt{m_i} \vec{x}_i,\\
    \vec{\xi}_i &=
                  \sqrt{\frac{\clustermass{i} m_{i+1}}{\clustermass{i+1}}}
                  \left(\frac{1}{\clustermass{i}}\sum_{j=1}^{i}
                  \sqrt{m_j}\vec{x}_j \right.\\
                &\left.-\frac{1}{\sqrt{m_{i+1}}}\vec{x}_{i+1}\right),
  \end{split}
\end{equation}    
where
\begin{equation}
  \begin{split}
    \clustermass{i} &= \sum_{j=1}^{i} m_j,\\
    \vec{x}_i &= \sqrt{\frac{m_i}{\mN}} \vec{r}_i,
  \end{split}
\end{equation}
with \(\vec{r}_i\) the particle coordinates and \mN{} the nucleon
mass, is particularly suitable for construction of the \HO{} basis
which is antisymmetric with respect to exchanges of nucleonic degrees
of freedom. In this set, the coordinate $\vec{\xi}_0$ is proportional
to the \CoM{} coordinate of the $A$-body system and the coordinates
$\vec{\xi}_i$, for $i=1, \ldots, A-1$, are proportional to the
relative position of baryon $i+1$ with respect to the \CoM{} of the
$i$-nucleon cluster. Once the single-particle coordinates and momenta
in the Hamiltonian~\eqref{eq:ncsm:h} are transformed into
coordinates~\eqref{eq:ncsm:jsetxi}, the kinetic energy term splits
into a part depending only on the \CoM{} coordinate $\vec{\xi}_0$ and
a part depending only on the intrinsic coordinates
$\{ \vec{\xi}_i \}_{i=1}^{A-1}$,
\begin{equation}
  \label{eq:ncsm:Tsplit}
  \sum_{i=1}^A
  \frac{-\hbar^2}{2m_i}\vec{\nabla}^{2}_{r_i} = \frac{-\hbar^2}{2 \mN}
  \sum_{i=0}^{A{-}1} \vec{\nabla}^{2}_{\xi_i}.
\end{equation}
This, together with translational invariance of $V_{\NN}$, $V_{\NNN}$,
and $V_{\YN}$, allows us to separate out the \CoM{} term and thus
decrease the number of degrees of freedom. As a result, the $A$-body
\HO{} basis states associated with Jacobi
coordinates~\eqref{eq:ncsm:jsetxi} with total angular momentum $J$ and
isospin $T$ can be constructed as
\begin{equation}
  \label{eq:ncsm:nonasbasis}
  \ket{ (\dotso(a_1,a_2)J_3T_3,a_3)J_4T_4,\dotsc,a_{\Amo})JT }.
\end{equation}
Here, $\ket{a_i} \equiv \ket{n_i(l_i s_i)j_i t_i}$ are \HO{} states
associated with coordinates $\vec{\xi}_i$, where $n_i$, $l_i$, $s_i$,
and $t_i$ are the radial, orbital, spin, and isospin quantum numbers,
respectively. The parentheses in~\eqref{eq:ncsm:nonasbasis} indicate
the coupling of angular momenta and isospins. The quantum numbers
$J_i$ and $T_i$ ($i=3,\dotsc,A$) are angular momentum and isospin
quantum numbers of $i$-baryon clusters (with $J_A\equiv J$ and
$T_A\equiv T$). Additionally, the \HO{} wave functions depend on a
single \HO{} frequency $\omega$ which is a free model-space parameter
in \YNCSM{} calculations.

Using Eq.~\eqref{eq:ncsm:Tsplit} it is, e.g., straightforward to
evaluate the matrix elements of the intrinsic kinetic energy between
the HO states~\eqref{eq:ncsm:nonasbasis} as
\begin{equation}
  \label{eq:ncsm:Tme}
  \left\langle
    \sum_{i=1}^{A}\frac{-\hbar^2}{2m_i}\vec{\nabla}^{2}_{r_i}  -
    \frac{-\hbar^2}{2 \mN}\vec{\nabla}^{2}_{\xi_0}\right\rangle
  = \sum_{i=1}^{A{-}1} \left\langle
    \frac{-\hbar^2}{2 \mN}\vec{\nabla}^{2}_{\xi_i}\right\rangle.
\end{equation}
Since the \HO{} potential transforms and separates into \CoM{} and
intrinsic parts in the same way as the kinetic energy
\begin{equation}
  \label{eq:ncsm:hotransf}
  \sum_{i=1}^A \frac{1}{2} m_i \omega^2\vec{r}_i^{\;2} = \sum_{i=0}^{A-1} \frac{1}{2} \mN \omega^2\vec{\xi}_i^{\;2},
\end{equation}
the matrix elements in Eq.~\eqref{eq:ncsm:Tme} can be simply evaluated
as
\begin{multline}
  \label{eq:ncsm:Ti}
  \left\langle\frac{-\hbar^2}{2
      \mN}\vec{\nabla}^{2}_{\xi_i}\right\rangle = \\
  \left\{
    \begin{array}{ll}
      \frac{\hw}{2}  \sqrt{(n'_i + 1)(n'_i + l_i + \frac{3}{2})} &\text{for~} n'_i = n_i - 1 \\
      \frac{\hw}{2} (2 n_i + l_i + \frac{3}{2}) &\text{for~} n'_i = n_i\\
      \frac{\hw}{2}  \sqrt{(n_i + 1)(n_i + l_i + \frac{3}{2})} &\text{for~} n'_i = n_i + 1.
    \end{array} \right.
\end{multline}
Here, the primed(non-primed) indices correspond to the initial(final)
state and the matrix elements are diagonal in all quantum numbers of
the state \eqref{eq:ncsm:nonasbasis} except for \(n_i\).
    
The set of Jacobi coordinates~\eqref{eq:ncsm:jsetxi} and the
associated antisymmetric basis states are; however, not convenient for
the evaluation of two-{} and three-body interaction matrix elements.
In order to evaluate the interaction matrix elements, different sets
of Jacobi coordinates are more suitable~\cite{Wirth:2017bpw}.

It is to be noted that the basis states~\eqref{eq:ncsm:nonasbasis} are
not antisymmetric with respect to exchanges of all $\Am1$ nucleons. To
construct a physical basis, fulfilling the Pauli exclusion principle,
the states~\eqref{eq:ncsm:nonasbasis} have to be antisymmetrized. In
\YNCSM{}, this is typically achieved by diagonalization of the
antisymmetrizer operator between the
states~\eqref{eq:ncsm:nonasbasis}. The matrix elements of the
antisymmetrizer, together with extensive discussion of the
antisymmetrization procedure, can be found in
references~\cite{Wirth:2017bpw} and~\cite{Navratil:1999pw}.

\YNCSM{} calculations are performed in a limited model space with
finite number of basis states to represent the Hamiltonian as a finite
matrix to be diagonalized. For a basis formed by the
states~\eqref{eq:ncsm:nonasbasis}, the size of the model space is
restricted by allowing only states with the total number of \HO{}
quanta restricted by
\begin{equation}
  \label{eq:ncsm:truncation}
  \sum_{i=1}^{A-1} (2n_i+l_i) \le \nm + N_0 \equiv \ntm,
\end{equation}
with $N_0$ the number of \HO{} quanta in the lowest state allowed by
symmetries. For \hnuc{3}{H}, \hnuc{4}{H}, \hnuc{4}{He}, and
\hnuc{5}{He} $N_0 = 0$. \YNCSM{} calculations are variational with respect
to the size of the model space and thus converge to exact results for
$\nm \rightarrow \infty$. 

\subsection{Infrared and ultraviolet scales}
\label{subsec:scales}
The finite size of the \YNCSM{} basis leads to model-space corrections
for observables, such as energies, computed in the \YNCSM{} basis.
This truncation of the oscillator space in terms of \nm{} and \hw{}
can be recast into associated \IR{} and \UV{} length-scale
cutoffs~\cite{Stetcu:2006ey, Jurgenson:2010wy, Coon:2012ab}. Only
recently, precise values of the \IR{} and \UV{} length scales were
identified for the \NCSM{} basis~\cite{Wendt2015}, which employs a
total energy truncation, as in~\eqref{eq:ncsm:truncation}. The key
insight in this case was that the finite oscillator space is, at low
energies, equivalent to confining particles by an infinite
hyper-radial well, the radius of which well then determines the \IR{}
length of the corresponding \NCSM{} basis.

Here we generalize the scheme in~\cite{Wendt2015} from nuclear \NCSM{}
to hypernuclear \YNCSM{} and extract the \IR{} length by equating the
lowest eigenvalues of the intrinsic kinetic energy operator in the
\YNCSM{} basis and in a \(D = 3(A-1)\) dimensional well with an
infinite wall at hyper-radius \lir.

Note that employing the mass-scaled relative Jacobi coordinates
\(\{\vec{\xi}_i\}_{i=1}^{A-1}\) from~\eqref{eq:ncsm:jsetxi} eliminates
the explicit dependence of the noninteracting Hamiltonian on the
unequal nucleon and hyperon masses and introduces a common (arbitrary)
mass scale \mN, see Eq.~\eqref{eq:ncsm:Tsplit}. It is thus convenient
to introduce hyper-spherical coordinates with the (squared)
hyper-radius \(\rho^2 = \sum_{i=1}^{A-1} \vec{\xi}_i^{\;2}\) and
hyper-radial states \(\ket{\rho\, G \bar\alpha}\), where \(G\) is the
grand angular momentum and \(\bar{\alpha}\) labels all other
partial-wave quantum numbers. This leads to the noninteracting
hyper-radial Schr\"{o}dinger equation
\begin{equation}
  \label{eq:scales:hreq}
  -\left( \partial^2_\rho - \LL(\LL+1)/\rho^2\right)
  \psi_{G}(\rho) = Q^2 \psi_{G}(\rho),
\end{equation}
where \(\LL = G + (D-3)/2\) and \(Q^2=2 m_N E / \hbar^2\) is the total
squared momentum. Imposing a Dirichlet boundary condition on
\(\psi_{G}\) at \(\rho = \lir\) completely determines the spectrum
\begin{equation}
  \label{eq:scales:Thrw}
  \left\{ Q^2_i \right\} = \left\{ \lir^{-2} X_{i,\LL} \,\vert\, i \in
    \mathbb{Z} \right\}
\end{equation}
by the hyper-radius \lir{}, the \(i\)th zero \(X_{i,\LL}\) of the
Bessel function \(J_{\LL+\frac{1}{2}}(Q\rho)\), and a minimal value of
\LL{} discussed below. 

To obtain the kinetic energy spectrum in the \YNCSM{} basis, we can
expand the three-dimensional \HO{} states~\eqref{eq:ncsm:nonasbasis}
in hyper-radial \HO{} basis states \(\ket{N G \bar{\alpha}}\). Here
\(N\) is the nodal quantum number of the hyper-radial coordinate
\(\rho\) and the transformation is diagonal in \(\ntm = 2N+G\). Matrix
elements of the \YNCSM{} kinetic energy operator between the
hyper-radial \HO{} states are diagonal in \(G\) and \(\bar{\alpha}\)
and can be evaluated as in Eq.~\eqref{eq:ncsm:Ti}. The
resulting spectrum can be written as
\begin{equation}
  \label{eq:scales:TNCSM}
  \left\{\frac{\hw}{2}T_{i,\LL}(\ntm)\right\},
\end{equation}
where \(T_{i,\LL}(\ntm)\) denotes the needed dimensionless
eigenvalues, and the smallest permitted eigenvalue is driven by the
smallest value of \LL{} allowed by the symmetries of the wave
function.

By equating the lowest eigenvalues in~\eqref{eq:scales:Thrw}
and~\eqref{eq:scales:TNCSM} we obtain the intrinsic IR length scale
\begin{equation}
  \label{eq:scales:lir}
  \lir = b\,\frac{X_{1,\mathcal{L}}}{\sqrt{T_{1,\mathcal{L}}}},
\end{equation}
where \(b = \sqrt{\hbar/(\mN \omega)}\) is the \HO{} length and
\begin{equation}
  \label{eq:scales:call}
  \mathcal{L} = G_{\mathrm{min}} + \frac{3(A-2)}{2}.
\end{equation}
The \(G_{\mathrm{min}}\) is the lowest value of the grand angular
momentum in the relative coordinate system determined by the sum of
relative orbital angular momenta which can couple with spins to yield
the ground-state parity and angular momentum \(J\)~\cite{Wendt2015}.
From the duality of the \HO{} Hamiltonian under the exchange of
position and momentum operators, the \UV{} scale of the \HO{} basis
can be identified~\cite{Konig:2014hma} as
\begin{equation}
  \label{eq:scales:luv}
  \luv = \frac{X_{1,\mathcal{L}}}{b \sqrt{T_{1,\mathcal{L}}}} = \frac{1}{b^2}\lir.
\end{equation}

Compared to the purely nuclear systems, the coupling to the
\(\Sigma\)-hypernuclear states and the incomplete antisymmetry of the
states only increases the degeneracy of the spectrum and modifies the
minimal value of \LL. See the Supplemental Material of
Ref.~\cite{Wendt2015} for tabulated values. It is to be noted,
however, that the effective hard-wall radius of \YNCSM{} at
\(\rho=\lir\) applies to the mass-scaled relative Jacobi
coordinates~\eqref{eq:ncsm:jsetxi} and not to the physical particle
separations. In fact, the physical separation between the hyperon
(particle $A$) and the $(A-1)$ nucleon cluster is
\begin{equation}
  \label{eq:rsep}
  \vec{r}_{\mathrm{sep}} = \frac{1}{A-1}\sum_{i\le A-1} \vec{r}_i -
  \vec{r}_A = \sqrt{\frac{\mN}{\mu_{A,A-1}}}\vec{\xi}_{A-1},
\end{equation}
where $\mu_{A,A-1} \equiv {\clustermass{A-1}m_A}/{\clustermass{A}}$ is
the reduced mass of the hyperon--$(A-1)$ nucleon system. The
associated separation momentum would be
\begin{equation}
  k_{\mathrm{sep}} = \frac{1}{\hbar} \sqrt{2\mu_{A,A-1} E_{\mathrm{sep}}},
  \label{eq:ksep}
\end{equation}
where $E_{\mathrm{sep}}$ is the separation energy.
  
\begin{figure}[tb]
  \begin{center}
    \includegraphics[width=0.95\columnwidth]{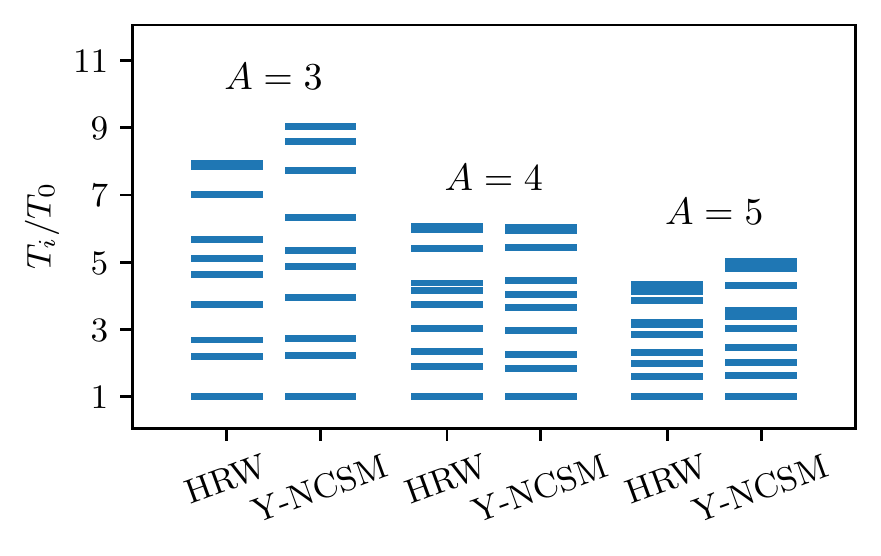}
  \end{center}
  \caption{Comparison of the discrete kinetic energy spectra for
    \(A=3, 4, 5\) hypernuclei in the \YNCSM{} and in the corresponding
    \(D=3(A-1)\) dimensional hyper-radial well (HRW). In each case,
    the spectrum is plotted in units of the smallest eigenvalue.}
  \label{fig:scales:HRW_YNCSM}
\end{figure}

As a verification of the relevant \IR{} length scale, we computed the
kinetic energy spectra for \(A=3,4,5\) hypernuclei in a \YNCSM{} basis
truncated at \(\nm = 10\) and the corresponding \(3(A-1)\)-dimensional
infinite hyper-radial wells. Their close similarity is demonstrated in
Fig.~\ref{fig:scales:HRW_YNCSM}. In each case, we plot the eigenvalues
\(T_i\) in units of the lowest eigenvalue \(T_0\) to remove the
proportionality of the entire spectrum to the inverse square of an
underlying length scale.

\subsection{Interactions}
\label{subsec:interactions}
A main focus of this work is to explore the importance of nuclear
physics uncertainties. For this purpose we employ the family of 42
different \nnlosim{} interactions~\cite{Carlsson:2015vda}
\footnote[1]{We have corrected the relation between $c_D$ in the
  one-pion exchange plus contact \NNN{} potential and the \LEC{}
  multiplying the contact axial-vector current, following the
  re-derivation by \citet{schiavilla:2018pc}.}
that are based on \chiEFT{} for nuclear systems up to \NNLO{}. At this
order, which is employed here, the nuclear interaction includes \NN{}
as well as \NNN{} forces. The 26 \LEC[s] of this interaction are
optimized to simultaneously reproduce \NN{} as well as \piN{}
scattering cross sections, the binding energies and charge radii of
\nuc{2,3}{H} and \nuc{3}{He}, the quadrupole moment of \nuc{2}{H} and
the Gamow-Teller matrix element associated with the beta-decay of
\nuc{3}{H}~\footnotemark[1].
Each \nnlosim{} potential is associated with one of seven
different regulator cutoffs $\Lam = 450, 475, \ldots, 575,
\unit[600]{MeV}$.
In addition, the database of experimental \NN{} scattering cross
sections used to constrain the respective interaction was also
varied. To be precise, it was truncated at six different maximum
scattering energies in the laboratory system
$\Tmax = 125, 158, \ldots, 257, \unit[290]{MeV}$.
A detailed description of the \nnlosim{} interactions and the
optimization protocol is given in Ref.~\cite{Carlsson:2015vda} .
The 42 different parametrizations of the nuclear interaction at this
\NNLO{} order give equally good descriptions of the relevant set of calibration
data in the nucleonic sector. Applying all of them within the ab
initio description of light hypernuclei will allow us to expose the
magnitude of systematic model uncertainties that stems from the
truncated \EFT{} description of the nuclear interaction.

For the \YN{} interaction we use the coupled-channel Bonn--J\"{u}lich
SU(3)-based \chiEFT{} model constructed at
\LO{}~\cite{Polinder:2006zh}. At \LO{}, \vyn{} consists of
pseudoscalar $\pi$, $K$ and $\eta$ meson exchanges, together with
baryon--baryon contact interaction terms. The meson--baryon coupling
constants and the form of the contact interaction is constrained by
the SU(3) flavor symmetry. The interaction is regularized in momentum
space by a smooth regulator,
\(f(p^\prime,p)=\exp(-(p^4 + p^{\prime 4})/\LamYN^4)\), with momentum
cutoff \(\LamYN\) ranging from 550 to \unit[700]{MeV}. Unless
otherwise specified we are using $\LamYN=\unit[600]{MeV}$. At \LO{},
there are five free parameters (\LEC[s]) which were determined from
the fits to the measured low-energy \YN{} scattering cross sections,
additionally conditioned by the existence of a bound \hnuc{3}{H} state
with \(J^\pi = \frac{1}{2}^+\)~\cite{Polinder:2006zh}.

The \nnlosim{} \NN{} and \loyn{} interactions are constructed in
particle basis, rather than isospin basis. To evaluate the
corresponding matrix elements between good-isospin \HO{} states we use
the prescription described in Ref.~\cite{Wirth:2017bpw}. This
procedure gives excellent agreement with particle-basis calculations,
as demonstrated in Ref.~\cite{Wirth:2017bpw}, where the difference
between total energies calculated in particle and isospin bases was
found to be a few keV for $A=3, 4$ hypernuclei. In the following, we
neglect this small contribution to the method uncertainty.

\section{Bayesian approach to infrared extrapolation}
\label{sec:bayesIR}
\subsection{Infrared extrapolation formalism}
\label{sec:irextrapol}
\begin{figure}[htb]
\begin{center}
  \includegraphics[width=0.95\columnwidth]{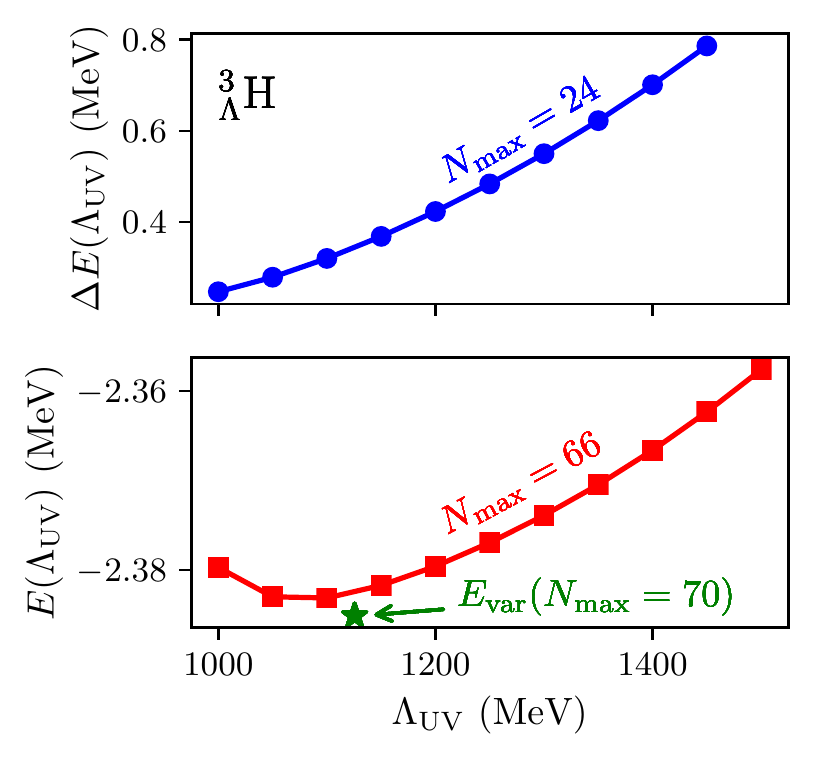}
\end{center}
\caption{The ground-state energy for \hnuc{3}{H} at fixed \nm{} as a
  function of the UV scale \luv{}. The results at fixed \luv{} are obtained using \GP{}
  interpolation {as described in the text}. The \UV{}-dependence of the extrapolation distance for
  smaller model spaces (upper panel) and the region of the variational minimum at very large model
  spaces (lower panel) indicates the $\luv \lesssim \unit[1200]{MeV}$ is a good choice
  for performing \IR{}-extrapolation.}
\label{fig:L3H_epol_LUV}
\end{figure}

Having established the IR and UV length scales of the \YNCSM{} basis
in Sec.~\ref{subsec:scales} we will employ an \IR{} extrapolation
formalism~\cite{furnstahl2012} to extract the infinite-model-space
energy eigenvalue, \Einf{}, from results, $E(\lir{}_i)$, computed at
truncated bases
\begin{equation}
  \label{eq:IRepol}
  E(\lir) = \Einf + a_0\exp{(-2 \kinf\lir)},
\end{equation}
where $\lir = \lir(A,\nm,\hw)$~\cite{Wendt2015}. In addition, there
will be \UV{} corrections to the computed energies. These errors can
be minimized by inferring the extra\-polation parameters using
computational results obtained at large and fixed
\luv{}~\cite{Forssen:2017wei}. Specifically, we will work at fixed
$\luv=\unit[1200]{MeV}$ which provides a good compromise between the
performance of reliable extrapolation and the minimization of \UV{}
corrections, as demonstrated in Fig.~\ref{fig:L3H_epol_LUV}. Note that
\hw{} can be tuned for fixed \nm{} to set different \UV{} scales. Here
we used \GP{} regression (with an RBF
kernel)~\cite{rasmussen2006gaussian} to interpolate in \HO{} frequency
for fixed \nm{}. The training data was composed of computed results at
even, integer values of $\hw \in [6,40]$~MeV. With this setup, and
using the \texttt{GPy} module~\cite{gpy2014}, we find that the
standard deviation for a predicted energy is less than \unit[0.5]{keV}
at any (interpolated) value of \hw{}.

Subleading \IR{} corrections to Eq.~\eqref{eq:IRepol}, here denoted
$\Delta E_{\IR,\NLO}$, are proportional to
$\exp{(-2 \kinf\lir)} / (\kinf\lir)$ as demonstrated in the two-body
case~\cite{furnstahl2014}. In many-body systems, further corrections
from additional separation channels are expected to be of the order of
$\exp{(-2k_\mathrm{sep}\lir)}$,~\cite{Forssen:2017wei}, where
$k_\mathrm{sep}$ is the relevant momentum scale. However, the latter
corrections will be suppressed in the present cases since the
$\Lambda$-separation threshold is much below other decay channels.

In practice, we will have computational results up to some largest
model space that translates into a maximum \IR{} length, \lirmax. We
define
\begin{equation}
  \label{eq:dlir}
  \dlir \equiv \lir - \lirmax,
\end{equation}
and replace the extrapolation factor $a_0$ that appear in
Eq.~\eqref{eq:IRepol} with a new parameter
\begin{equation}
  \label{eq:dEinf}
  \dEinf \equiv a_0  \exp{ (-2\kinf \lirmax )}.
\end{equation}
This transformed parameter corresponds to the size of the \LO{} \IR{} energy correction for our largest
model-space result. Furthermore, we introduce a random variable
 \epsNLO{} that is expected to be of natural size and that will provide
a stochastic model for the \NLO{} energy correction. Consequently, we
have the extrapolation model
\begin{equation}
  \begin{split}
    E(\lir) = \Einf + \dEinf \exp{(-2 \kinf\dlir)} \\
    \times \left( 1 + \frac{\epsNLO}{\kinf(\lirmax+\dlir)} \right).
    \label{eq:statIRepol}
  \end{split}
\end{equation}
We also note that the \kinf{} parameter is related to the lowest
separation energy threshold of the many-body system~\cite{Forssen:2017wei}. 
From the asymptotic form of the wave function
\begin{equation}
  \exp{\left(-k_{\mathrm{sep}}r_{\mathrm{sep}}\right)} =
  \exp{\left(-\kinf \xi_{A-1}\right)},
\end{equation}
and assuming that the lowest separation energy is well below the
second lowest one, we get that \kinf{} from the fit should be related to \(k_{\mathrm{sep}}\) by
\begin{equation}
  \kinf = \sqrt{\frac{\mN}{\mu_{A,A-1}}} k_{\mathrm{sep}}
  = \frac{1}{\hbar} \sqrt{2\mN E_{\mathrm{sep}}},
  \label{eq:kinf}
\end{equation}
where we have used the relations in Eqs.~\eqref{eq:rsep} and
\eqref{eq:ksep}. In the following we will use \ksepexp{} (\ksepth{})
to denote the momentum associated with the lowest experimental
(theoretical) separation threshold energy while \kinf{} will always be
the fit parameter. Higher-order corrections, such as the effects of
other separation channels, will in practice lead to
$\kinf \gtrsim \ksepth{}$.

\subsection{Bayesian inference}
The application of the extrapolation model~\eqref{eq:statIRepol}
becomes an inference problem that we tackle using a Bayesian approach.
Our \YNCSM{} computations for a specific Hamiltonian provide a set of
$N$ energies
$\mathcal{D} = \{ E(\lir{}_{,1}), E(\lir{}_{,2}), \ldots,
E(\lirmax)\}$ obtained in different model spaces (corresponding to
\IR{} cutoffs $\lir{}_{,1} < \lir{}_{,2} < \ldots < \lirmax$). We will
assume that the corresponding vector of \NLO{} errors, $\vec \epsNLO$,
is normally distributed $\pr(\vec\epsNLO) = \mathcal{N}(0,\Sigma)$.
The specification of the covariance matrix $\Sigma$ will involve
additional model parameters. With $\vec{\alpha}$ the vector of model
parameters (\IR{} extrapolation model plus statistical error model)
this assumption implies that the data likelihood
$\pr( \mathcal{D} \given \vec{\alpha} )$ becomes a normal
distribution. As mentioned, the stochastic variable(s) \epsNLO{} are
expected to be of natural size, which we express as
$\text{Cov}\left( \epsNLO{}(\lir{}_{,i}),
  \epsNLO{}(\lir{}_{,i})\right) = \epsbar{}^2$ with \epsbar{} of order
unity. Furthermore, we expect that the \NLO{} correction is a rather
smooth function of \lir{}, which we translate into an assumption of
positive correlations
$0 < \text{Cov}\left( \epsNLO{}(\lir{}_{,i}),
  \epsNLO{}(\lir{}_{,i+1})\right) / \epsbar{}^2< 1$. We adopt a rather
simple model for this correlation structure that involves a single
unknown parameter $\rho$ describing the correlation coefficient
between two subsequent model space results. The expected decay of the
correlation strength with increasing \IR{} distance is here
implemented by assigning a Toeplitz structure to the correlation
matrix
\begin{equation}
  C = \begin{pmatrix}
    1 & \rho & \rho^2 & \cdots & \rho^{N-1}\\
    \rho & 1 & \rho &  \cdots & \rho^{N-2} \\
    \rho^2 & \rho & 1 & & \vdots \\
    \vdots & \vdots & & \ddots \\
    \rho^{N-1} & \rho^{N-2}& \cdots & & 1
    \end{pmatrix}.
\label{eq:toeplitz}
\end{equation}
We note that the $N \times N$ covariance matrix $\Sigma$ is determined
by the two parameters \epsbar{} and $\rho$.

The calibration of all model parameters
$\vec{\alpha} = \{ \Einf, \dEinf, \kinf, \epsbar, \rho\}$ is an
inference problem that we approach using Bayes formula
\begin{equation}
  \pr( \vec{\alpha} \given \mathcal{D} ) \propto \pr (  \mathcal{D}
  \given \vec{\alpha} ) \pr (\vec{\alpha} ).
  \label{eq:bayes}
\end{equation}
The \PDF{} on the left-hand side of Eq.~\eqref{eq:bayes} is the posterior which is
proportional to the product of the 
likelihood \PDF{} for the data (conditional on model parameters) and the prior \PDF{}
for the parameters. We will
usually refer to these \PDF[s] as just posterior, likelihood, and prior, respectively.

The first step is then to formulate our prior beliefs for the model parameters.

\subsection{Priors for model parameters}
First we note that the original parametrization of the extrapolation
formula~\eqref{eq:IRepol} is characterized by a very strong
correlation between model parameters $a_0$ and
\kinf~\cite{Forssen:2017wei}. Using the transformed parameters,
\dEinf{} and \kinf{}, we expect to find more independent constraints.
In fact, we will make no prior assumption of correlations and will
assign full factorization
\begin{equation}
  \pr(\vec{\alpha}) = \pr(\Einf) \pr(\dEinf) \pr(\kinf) \pr(\epsbar) \pr(\rho).
\end{equation}
Our choices of priors for \Einf{}, \dEinf{}, and \kinf{} are guided by
studies of the \nm{} and \hw{} dependence of our results. At this
stage we focus on \HO{} frequencies close to the variational minimum
rather than the large values that are used for the \IR{} extrapolation
model inference. The key output of this pre-study is
$(\Evar,\dEinfmax)$ where \Evar{} is the (approximate) variational
minimum and \dEinfmax{} is a very generous estimate for the maximum
extrapolation distance. Furthermore, we use \Evar{} to compute a very
rough estimate of the separation energy and the associated momentum
scale $\ksep{th,var} = \sqrt{ 2 \Esep{th,var} \mN} / \hbar$. We can
now specify conservative prior bounds implemented via uniform
distributions
\begin{align}
  \pr( \Einf{} ) &= \mathcal{U} ( \Evar - \dEinfmax, \Evar{}+ \delta\Evar{} ), \\
  \pr( \dEinf{} ) &= \mathcal{U} ( 0, \dEinfmax), \\
  \pr( \kinf{} ) &= \mathcal{U} (\frac{\ksep{th,var}}{3}, 3\ksep{th,var}  ), 
  \label{eq:prior_epol}
\end{align}
where $\delta\Evar = \unit[0.1]{MeV}$ is an extra flexibility to allow for
possible \UV{} errors at low \HO{} frequencies.

Concerning the error model parameters \epsbar{} and $\rho$ we assign
\begin{align}
  \pr( \rho{} ) &= \mathcal{U} ( 0.01, 0.99), \\
  \pr( \epsbar{} ) &= f(\alpha=1.5, \beta=1.0), 
  \label{eq:prior_error}
\end{align}
where the former encapsulates our expectation that correlations will
be positive but of unknown strength, and the latter is a weakly
informative inverse gamma distribution. This particular
parametrization gives a main strength for natural values (the
probability mass for $\epsbar < 2.0$ is 0.8) while still allowing for
larger values via a significant tail.

\subsection{MCMC sampling}
\label{sec:mcmc}
\begin{figure}[tb]
  \begin{center}
    \includegraphics[width=0.95\columnwidth]{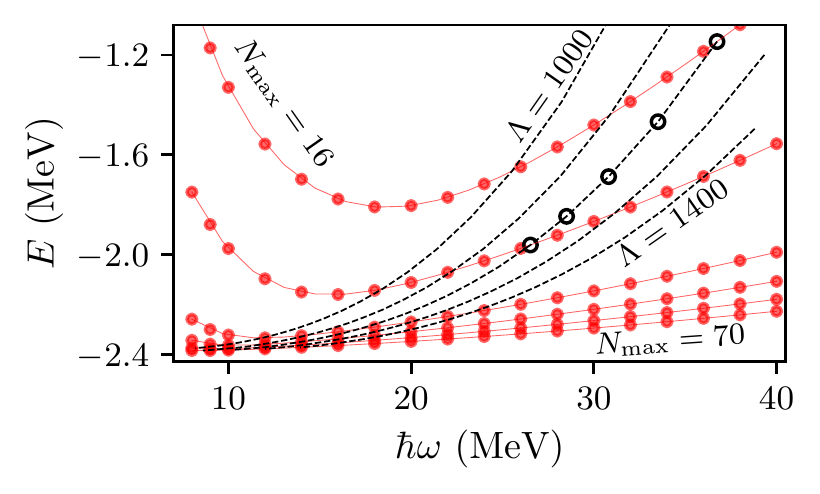}
  \end{center}
  \caption{The ground-state energy of \hnuc{3}{H} as a function of the
    \HO{} frequency and \YNCSM{} model space \nm{}. Results obtained
    with \nnlosimstd{} at a subset of
    fixed $\nm=\{16,24,40,50,60,70\}$ are indicated by filled symbols
    connected by thin, solid lines. Dashed lines indicate fixed \UV{}
    scale $\luv = \{1000,1100,1200,1300,1400\}$~MeV. The benchmark \IR{}
    extrapolation shown in Fig.~\ref{fig:L3H_epol_bayesian} is
    performed with the \GP{}-interpolated data indicated by open
    circles.}
  \label{fig:L3H_hw}
\end{figure}
Having specified the likelihood for the calibration data and the
priors for the parameters we are in a position to collect samples from
the posterior \PDF{}~\eqref{eq:bayes} using \MCMC{} methods. Here we
use the affine invariant \MCMC{} ensemble sampler
\texttt{emcee}~\cite{Foreman:2013} using up to {100} walkers with
{50,000} iterations per walker following {5,000} warmup steps.

\begin{figure}[tb]
  \begin{center}
    \includegraphics[width=0.95\columnwidth]{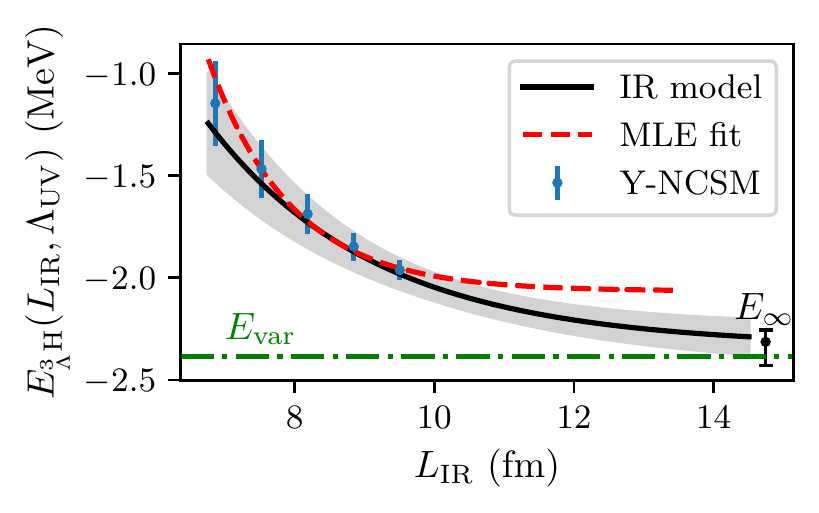}
  \end{center}
  \caption{Bayesian \IR{} extrapolation of the ground-state energy for
    \hnuc{3}{H} with NLO errors. A truncated data set ($\nm = 16-24$) with fixed
    $\luv=\unit[1200]{MeV}$ is used for the extrapolation. The posterior IR
    model prediction is represented by the band that shows the median
    (black line) and the 68\% credible region. The median and 68\%
    credible interval for the \Einf{} parameter is
    shown by the black marker with error bar.
    The green, dashed line
    indicates the variational minimum obtained at $\nm=70$,
    $\hw=\unit[9]{MeV}$~\cite{Htun:2021jnu}. The red line shows a maximum-likelihood fit
    to the data inversely weighted by the \NLO{} errors.}
  \label{fig:L3H_epol_bayesian}
\end{figure}

\begin{figure*}[htb]
  \begin{center}
    \includegraphics[width=0.85\textwidth]{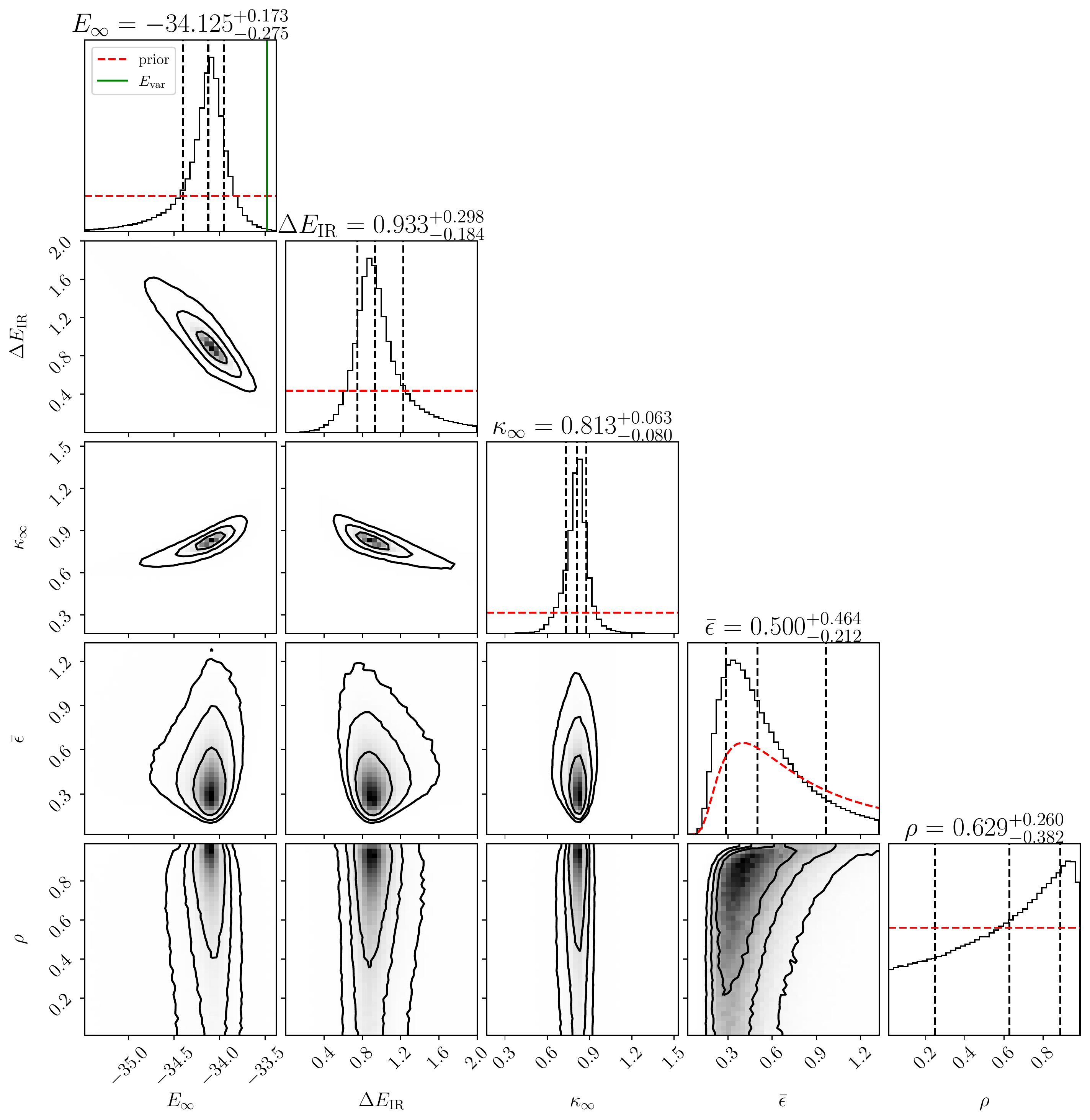}
  \end{center}
  \caption{Bayesian extrapolation results for \hnuc{5}{He} with the
    \nnlosimstd{} nuclear interaction, and the Bonn--J\"{u}lich \LO{}
    \YN{} interaction, computed at $\nm=4-10$ and fixed
    $\luv=\unit[1200]{MeV}$. Energies are in MeV and \kinf{} in
    fm$^{-1}$. The median and the 68\% equal-tail credible interval
    are indicated with vertical black, dashed lines for the marginal
    distributions on the diagonal. The prior distributions are shown
    by red, dashed lines while the variational estimate for the energy
    is shown by a green, solid line n the upper left panel. The
    contour lines for the bivariate distributions indicate {39\%,
      68\%, 86\%} probability masses.}
  \label{fig:L5He_epol_bayesian}
\end{figure*}

We check the performance and consistency of our Bayesian approach to
\IR{} extrapolation by first studying the \hnuc{3}{H} system. Here we
can perform calculations in very large model spaces up to $\nm = 70$
and we have extracted a well-converged variational minimum
$\Evar = \unit[-2.385]{MeV}$ for the \nnlosimstd{}
interaction~\cite{Htun:2021jnu}. For testing purposes, we limit the
\IR{} extrapolation data to $\nm \le 24$. We determine the prior by
studying the behavior near the variational minimum, and we select
calibration data for the likelihood that has fixed
$\luv=\unit[1200]{MeV}$, see Fig.~\ref{fig:L3H_hw}. This implies an
extrapolation distance of almost \unit[500]{keV} as seen in
Fig.~\ref{fig:L3H_epol_bayesian} and we find that a simple \MLE{} fit
to the data (with \NLO{} errors as inverse weights) fails to capture
the large-\lir{} behavior (red, dashed line) and severely
underestimates the converged binding energy. In contrast, the Bayesian
approach provides extrapolation model samples (gray band) that are
consistent with the calibration data. The median (inferred 68\%
credible interval) for \Einf{} is $-2.31$ ($[-2.43,-2.26]$)~MeV (black
symbol with error bar), which encompasses the variational minimum
(green, dash-dotted line).

For \hnuc{4}{H} and \hnuc{4}{He} $0^+$ states we are able to reach
$\nm=20$ and the extrapolation distance is less than \unit[100]{keV}.
As will be shown in Sec.~\ref{sec:results} the extrapolation
uncertainty is just $\sim \unit[10]{keV}$. We also compute the $1^+$
states in these $A=4$ hypernuclei, but are then limited to $\nm=16$.
The extrapolation distance is around \unit[400]{keV} and, as we will
show, the associated uncertainty becomes $\sim \unit[100]{keV}$.

The most challenging calculation is for \hnuc{5}{He} with computations
limited to $\nm \le 10$ and results not yet close to converged. The
variational minimum in this model space is
$\Evar = \unit[-33.48]{MeV}$ while the \IR{} model calibration data
extends down to \unit[-33.19]{MeV}. The parameter posterior for the
Bayesian extrapolation analysis is shown in
Fig.~\ref{fig:L5He_epol_bayesian}. We note the slightly asymmetric
mode for \Einf{}, with a long negative tail, and its anti-correlation
with \dEinf{}, which is quite expected. The magnitude of the \NLO{}
error is smaller than the prior assumption. The correlation
coefficient is not that well constrained, but has its main support for
rather strong correlations. The bivariate distribution
$\pr( \epsbar, \rho \given \mathcal{D})$ indicates that strongly
correlated errors allow for a larger \NLO{} correction which is also
expected since the summed penalty in the likelihood decreases with
increasing correlation.

The priors and posteriors for all extrapolation parameters for the
ground-state energies of \hnuc{4}{H} and \hnuc{4,5}{He} and the $1^+$
excited states of \hnuc{4}{H,He} are summarized in
Table~\ref{tab:hyperpars} in the Appendix.

\section{Nuclear physics uncertainties}
\label{sec:results}

\subsection{Binding and \(\Lambda\) separation energies}
\label{subsec:totalE}
Energy levels of light hypernuclei are experimentally accessible
observables that are sensitive to details of the underlying \YN{} and
nuclear interactions. Yet, one can naively expect that calculated
\(\Lambda\) separation energies---obtained as the differences of the
binding energies of hypernuclei and their core nuclei---should be
insensitive to the choice of nuclear interaction. In fact, such a
rather weak residual dependence of \(\Lambda\) separation energies in
\(A=3,4\) hypernuclei was found already in Faddeev
calculations~\cite{Nogga:2001ef} using a limited set of
phenomenological \NN{} interactions and, more recently, also in
\NCSM{} calculations using \chiEFT{} \NN{} interaction
models~\cite{Le:2020zdu}. However, our initial analysis for
\hnuc{3}{H} in Ref.~\cite{Htun:2021jnu} indicated that this dependence
may be significantly larger.

In this work we therefore carry a comprehensive systematic study of
the variation of the binding and \(\Lambda\) separation energies of
light hypernuclei resulting from the uncertainties in \NN{} and \NNN{}
potentials. In order to quantify this model uncertainty we employed
the \nnlosim{} family of nuclear interactions~\cite{Carlsson:2015vda},
specifically designed for such
tasks~\cite{Gazda:2016mrp,Acharya:2016kfl,Htun:2021jnu}. In addition,
we make use of the Bayesian approach to \IR{} extrapolation from
Sec.~\ref{sec:bayesIR} to determine the accompanying method
uncertainty, associated with the solution of the many-body problem in
a truncated \YNCSM{} model space.
\begin{figure}[tb]
  \begin{center}
    \includegraphics[width=0.95\columnwidth]{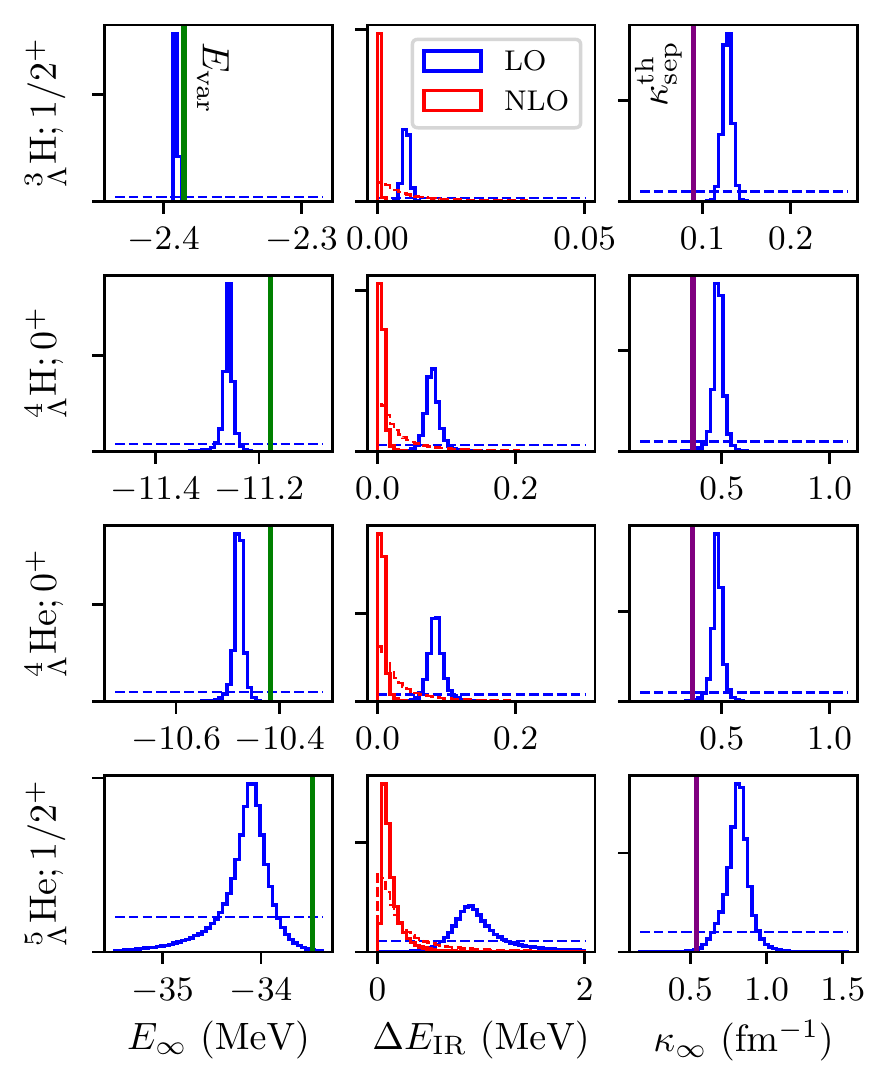}
  \end{center}
  \caption{Posterior \PDF[s] for the extrapolated energies (first
    column) as well as \LO{} and \NLO{} \IR{} corrections (second
    column) and \kinf{} (third column) with the \nnlosimstd{}
    interaction for \hnuc{3,4}{H} and \hnuc{4,5}{He}. The variational
    minimum energies, \Evar{}, are shown by green, vertical lines in
    the first column while the theoretical separation momenta,
    \ksepth{}, are shown by purple, vertical lines in the final
    column. The priors are indicated by dashed, blue lines.}
  \label{fig:epol_bayesian}
\end{figure}
Results of this technique for the ground states of \hnuc{3,4}{H} and
\hnuc{4,5}{He} are represented in Fig.~\ref{fig:epol_bayesian} by
posterior \PDF[s] of the extrapolated binding energies (first column)
as well as \LO{} and \NLO{} \IR{} corrections (second column) and
\kinf{} (third column) using a single \nnlosim{} nuclear interaction
with \(\Lam = \unit[500]{MeV}\) and \(\Tmax = \unit[290]{MeV}\). We
note that the precision of the inferred energy, that is represented by
the width of the \PDF{} for \Einf{}, depends critically on the
extrapolation distance, i.e., the magnitudes of the \IR{} corrections.
We also confirm the relation between the \kinf{} fit parameter and the
(theoretical) separation momentum, \ksepth{}, (here obtained with the
median value for \Esepth{}) as discussed in Sec.~\ref{sec:irextrapol}.
\begin{figure}[tb]
  \begin{center}
    \includegraphics[width=0.95\columnwidth]{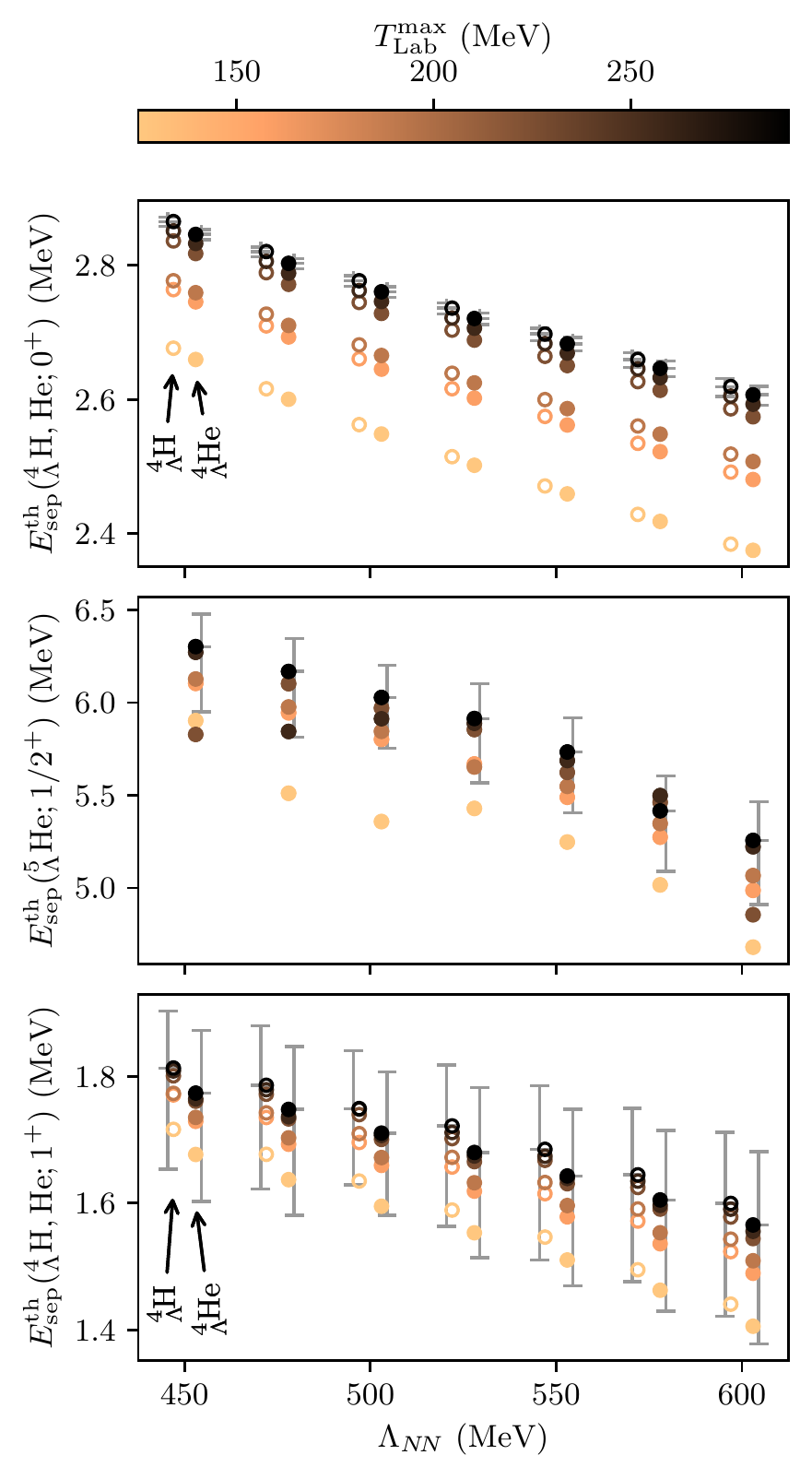}
  \end{center}
  \caption{Separation energies for \hnuc{4}{H} (open circles) and
    \hnuc{4,5}{He} (filled circles) ground-{} and excited states for the family of
    \nnlosim{} interactions. The data corresponds to the medians from
    the Bayesian \IR{} extrapolation. The 68\% credible region from the
    \IR{} extrapolation is indicated by the error bar for the seven 
    $\Tmax=\unit[290]{MeV}$ interactions in the \nnlosim{} family.}
  \label{fig:L45_esep}
\end{figure}
All inferred parameters of the \IR{} extrapolation are summarized in
Table~\ref{tab:hyperpars} in the Appendix, including also the $1^+$
excited states in \hnuc{4}{H,He} and prior distributions. There, we
characterize the posteriors of the parameters by the median value of
the distribution plus the 68\% and 95\% credible intervals. The half
width of the 68\% credible region, {quantifying} the extrapolation
(method) uncertainty, ranges from \unit[1(10)]{keV} for the ground
states of \hnuc{3}{H} (\hnuc{4}{H,He}), to \unit[100]{keV} for the
excited $1^+$ states in \hnuc{4}{H,He} and \unit[200]{keV} in
\hnuc{5}{He}. These numbers are obtained with the \nnlosimstd{}
interaction. We note (see Fig.~\ref{fig:L45_esep}) that the
extrapolation uncertainty becomes larger with increasing \Lam{}
regulator cutoff. The main limitation in precision originates in the
computation restriction, constraining the feasible size of the
\YNCSM{} model space truncation \nm{}.

Having assessed the method error, we apply the \IR{} extrapolation for
all nuclear interactions in the \nnlosim{} family. The $\Lambda$
separation energies for all 42 interactions are shown in
Fig.~\ref{fig:L45_esep} for \hnuc{4}{H} (open circles) and
\hnuc{4,5}{He} (filled circles) ground-{} and excited states. The data
corresponds to the medians from the Bayesian \IR{} extrapolation while
the error bars for the seven $\Tmax=\unit[290]{MeV}$ \nnlosim{}
interactions with different \Lam{} indicate the 68\% credible region
from the \IR{} extrapolation. We use the variance,
$\sigma^2\left( \mathrm{\nnlosim} \right)$, of predictions for
\Esep{th} obtained with the full \nnlosim{} family to quantify the
uncertainty connected to the choice of nuclear Hamiltonian. The
resulting model uncertainties
\begin{equation}
  \label{eq:sigmamodel}
  \sigma_\mathrm{model} \equiv \left[ \sigma^2 \left( \mathrm{\nnlosim} \right) \right]^{1/2},
\end{equation}
as well as the method uncertainties from the \IR{} extrapolations, are
summarized in Table~\ref{tab:YN_obs} for all hypernuclear states
studied in this work. The model uncertainties remain {significant}
despite the fact that the method uncertainties become comparable (or
even larger) in magnitude for \hnuc{5}{He} and the \(1^+\) excited
states of \hnuc{4}{H,He}. It should be noted that each interaction in
the \nnlosim{} family gives slightly different binding energy for
\nuc{4}{He}~\cite{Carlsson:2015vda}, which is taken into account when
computing the separation energies. For example, the \nnlosimstd{}
interaction gives
$E(\nuc{4}{He})=\unit[-28.10]{MeV}$~\footnotemark[1]. We also note
that a direct comparison of \(\sigma_{\mathrm{model}}\) with previous
studies, such as in Refs.~\cite{Nogga:2001ef, Le:2020zdu}, could be
misleading. Usually reported is the total spread of \(\Lambda\)
separation energies obtained by using a very limited set of nuclear
interactions. Moreover, \NCSM{} calculations might additionally suffer
from an undesired dependence on the flow parameter of the \SRG{}
transformation applied in order to speed up the convergence. In this
work we don't use SRG transformations but rather make an effort to
quantify the method uncertainty associated with the convergence.
\begin{table}[b]
  \caption{Separation energies (in MeV) for light hypernuclei.
    Experimental data is compared with the ab~initio result of this
    work (\Esepth{}) obtained with \nnlosimstd{}. The 68\% credible
    interval (CI), shown as distances from the median in the next to
    final column, corresponds to the uncertainty of the extrapolation
    procedure. The model discrepancy in the final column
    ($\sigma_\mathrm{model}$) is an estimate of the
    nuclear-interaction uncertainty. The model discrepancy is here
    quantified by the standard deviation of predictions obtained with
    the full \nnlosim{} family, see Eq.~\eqref{eq:sigmamodel}.}
  \label{tab:YN_obs}
  \begin{ruledtabular}
    \begin{tabular}{c d{3.6} c | d{3.2} c d{3.2}}
      \multicolumn{1}{c}{System}
      & \multicolumn{1}{c}{\Esepexp}
      & \multicolumn{1}{c}{Ref.}
      & \multicolumn{3}{c}{\Esepth} 
      \\
      & &
      & \multicolumn{1}{c}{median}
      & \multicolumn{1}{c}{68\% CI$_\mathrm{method}$}
        & \multicolumn{1}{c}{$\sigma_\mathrm{model}$}
      \\
      \hline
      $\hnuc{3}{H}$
      & \unit[0.165(44)]{} & \cite{Eckert:2022dyz}
      & \unit[0.166]{} & $[-0.001,+0.001]$  & \unit[0.02]{} \\[1ex]
      $\hnuc{4}{H}$ & \unit[2.157(77)]{} &
                                           \cite{A1:2016nfu} & \unit[2.78]{} & $[-0.01,+0.01]$ & \unit[0.08]{} \\[1ex]
      $\hnuc{4}{He}$ & \unit[2.39(3)]{} &
                                          \cite{Davis2005} & \unit[2.76]{} & $[-0.01,+0.01]$ & \unit[0.08]{} \\[1ex]
      $\hnuc{5}{He}$ & \unit[3.12(2)]{} &
                                          \cite{Davis2005} & \unit[6.03]{} & $[-0.28,+0.18]$ & \unit[0.36]{} \\[1ex]
      $\hnuc{4}{H}; 1^+$ &  \unit[1.067(80)]{} &
                                                 \cite{A1:2016nfu} & \unit[1.75]{} & $[-0.12,+0.10]$ &  \unit[0.07]{}\\[1ex]
      $\hnuc{4}{He}; 1^+$ &  \unit[0.984(50)]{} &
                                                  \cite{Yamamoto2015} & \unit[1.71]{} & $[-0.13,+0.10]$ & \unit[0.07]{}
    \end{tabular}
  \end{ruledtabular}
\end{table}

Finally, it should be stressed that all many-body computations
discussed so far have been performed with fixed \YN{} regulator cutoff
$\LamYN = \unit[600]{MeV}$. The \loyn{} interaction is known to result
in a noticeable cutoff dependence of separation energies in light
\(A=3,4\) hypernuclei~\cite{Nogga:2013pwa}, as well as heavier systems
using \SRG{}-evolved \YN{}
interactions~\cite{Wirth:2017bpw,Le:2020zdu}. We find that the
\hnuc{5}{He} binding energy, in particular, is sensitive to the choice
of \(\LamYN\). As shown in Table~\ref{tab:L5He_e_YN}, larger values of
\(\LamYN\) seem to give a better agreement with the experimental
separation energy shown in Table~\ref{tab:YN_obs}, but are also
associated with a larger extrapolation uncertainty. A more complete
study of this sensitivity is left for future work and should possibly
also include higher-order descriptions of the \YN{} interaction. We
note that most few-body calculations employing various \YN{}
interaction models that reproduce ground-state \(\Lambda\) separation
energies of lighter, \(A \le 4\), hypernuclei yield too large
\hnuc{5}{He} \(\Lambda\) separation energy. See
Ref.~\cite{Contessi:2018qnz} for an overview of available calculations
and a discussion of this issue within pionless EFT and the
implications of the strength of three-body \YNN{} interactions for
neutron-star matter. The uncertainty quantification presented here
should be relevant in the resolution of this puzzle.

\begin{table}[tb]
  \caption{Separation energies for \hnuc{5}{He} with different
    regulator cutoffs for the \LO{} \YN{} interaction. The nuclear
    interaction (\NN{}+\NNN{}) is fixed as \nnlosimstd{}. The Bayesian
    extrapolations are performed at fixed $\luv = \unit[1200]{MeV}$
    including \NLO{} \IR{} errors as described in
    Sec.~\ref{sec:bayesIR}. Median values from the sampled \PDF[s] are
    shown for the extrapolation distance \dEinf{} and \Esepth{}.
    Furthermore, the 68\% and 95\% credibility intervals for \Esepth{}
    are shown in the last two columns as distances from the median.
    All values in MeV.}
  \label{tab:L5He_e_YN}
  \begin{ruledtabular}
    \begin{tabular}{c d{3.2} | d{3.2} c c}
      \multicolumn{1}{c}{$\LamYN$}
      & \multicolumn{1}{c}{\dEinf}
      & \multicolumn{3}{c}{\Esepth} \\
      & & \multicolumn{1}{c}{median}
      & \multicolumn{1}{c}{68\% CI$_\mathrm{method}$}
        & \multicolumn{1}{c}{95\% CI$_\mathrm{method}$}
      \\
      \hline
      550 & 0.72 & 7.33 & [-0.20, +0.13] & [-0.55, +0.30]\\
      600 & 0.93 & 6.03 & [-0.28, +0.18] & [-0.75, +0.41]\\
      650 & 1.23 & 4.79 & [-0.37, +0.25] & [-0.94, +0.56]\\
      700 & 1.56 & 3.82 & [-0.46, +0.33] & [-1.11, +0.69]\\
    \end{tabular}
  \end{ruledtabular}
\end{table}

\subsection{Charge symmetry breaking}
\label{subsec:CSB}

Unlike in nuclei, the amount of \CSB{} in hypernuclei is substantial
as it originates in the strong \YN{} interaction. While heavily
suppressed in \hnuc{3}{H} with \(T=0\), it manifests itself in the
\(\Lambda\) separation energy differences of the \(A=4\)
\hnuc{4}{H,He} mirror hypernuclei
\begin{multline}
  \label{eq:csb:DeltaEsep}
  \DEsep{J^\pi}{} \\ =
  \Esep[\hnuc{4}{He};J^\pi]{} -
  \Esep[\hnuc{4}{H};J^\pi]{}.
\end{multline}
The large \(\DEsep{0^+}{exp} = \unit[0.233(92)]{MeV}\) and negligible
\(\DEsep{1^+}{exp} = \unit[-0.083(94)]{MeV}\) \CSB{} effects in
\hnuc{4}{H,He} ground and excited states, respectively, were
reaffirmed recently by precision
measurements~\cite{A1hypernuclear:2015rzq,J-PARCE13:2015uwb}. These
observations provide unique information on the charge dependence of
the \YN{} interaction. Given that low-energy \(\Lambda p\) cross
sections are poorly known and \(\Lambda n\) scattering data do not
exist, the so far available chiral \YN{} interaction
models~\cite{Polinder:2006zh,Haidenbauer2013} have assumed isospin
symmetry. Leading \CSB{} terms were only recently incorporated into
the chiral NLO~\YN{} interaction and constrained to reproduce the
measured \(\Lambda\) separation energy differences in
\hnuc{4}{H,He}~\cite{Haidenbauer:2021wld}.

\begin{figure}[t]
  \centering
  \includegraphics[width=0.95\columnwidth]{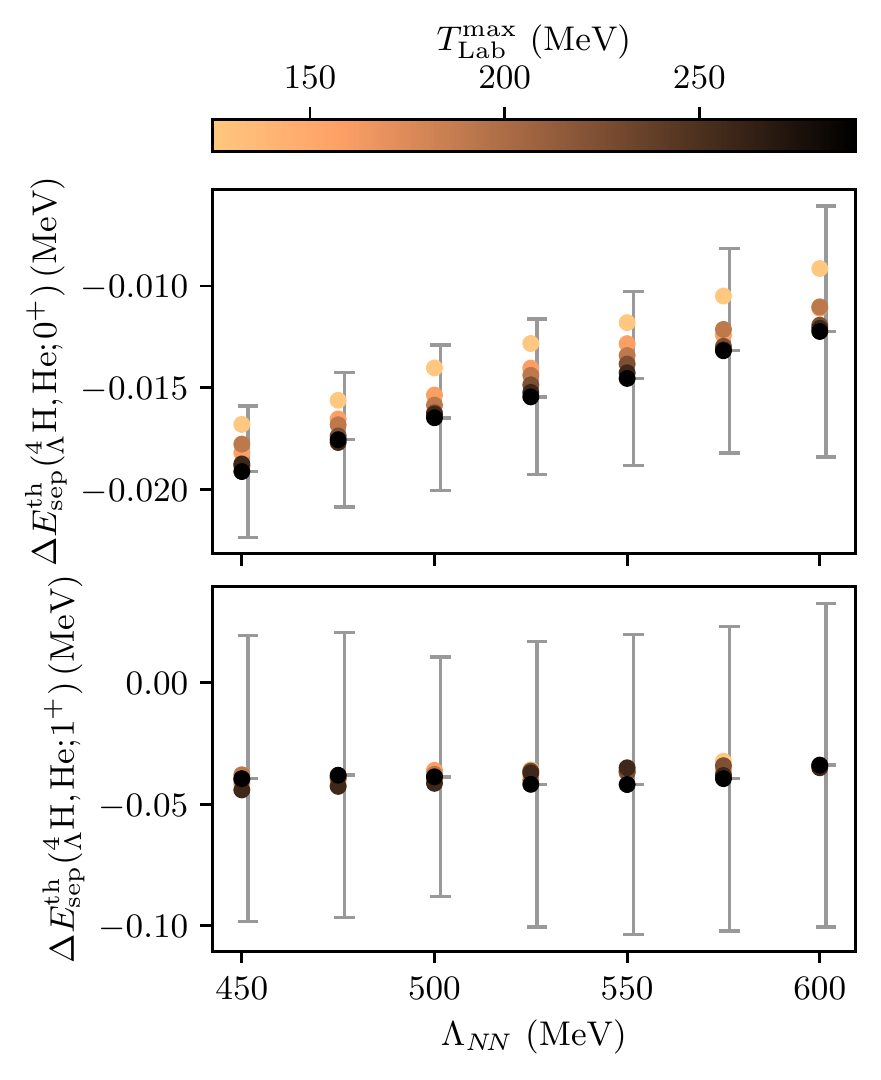}
  \caption{Difference of ground-{} and excited-state \(\Lambda\)
    separation energies in \hnuc{4}{He} and \hnuc{4}{H} for the family
    of \nnlosim{} interactions, calculated using the charge-symmetric
    LO~\YN(\LamYN=\unit[600]{MeV}) interaction. The data corresponds
    to the medians from the Bayesian IR extrapolation. The error bars,
    shown for the seven $\Tmax=\unit[290]{MeV}$ interactions in the
    \nnlosim{} family, represent a conservative estimate of the
    uncertainty obtained from the uncertainties of the separation
    energies assuming correlated method errors (correlation
    coefficient $r=0.9$; see text for details).}
  \label{fig:CSB}
\end{figure}

Here we quantify the precision of theoretical predictions for
\(\DEsep{J^\pi}{th}\), considering both model and method
uncertainties. This effort reveals the potential of this observable to
constrain the charge dependence of the \YN{} interaction. The
sensitivity of the \(\Lambda\) separation energy differences on the
nuclear interaction model in \(0^+\) and \(1^+\) states of
\hnuc{4}{H,He} is shown in Fig.~\ref{fig:CSB}, using the
charge-symmetric LO~\YN(\LamYN=\unit[600]{MeV}) interaction. The
\DEsep{J^\pi}{th} were obtained using the median values of
extrapolated \(\Lambda\) separation energies,
\(\Esepth{(\hnuc{4}{H,He};{J^\pi})}\), for all 42 interactions in the
\nnlosim{} family. The small residual \CSB{} splittings,
\(\DEsep{0^+}{th} \approx \unit[-0.015]{MeV}\) and
\(\DEsep{1^+}{th} \approx \unit[-0.038]{MeV}\), are due to the
increased Coulomb repulsion in \hnuc{4}{He} compared to its
\nuc{3}{He} core~\cite{Bodmer:1985km} and the \(\Sigma N\)
intermediate-state mass differences in kinetic energy
terms~\cite{Nogga:2001ef,Gal:2015bfa}. These results are consistent
with previous calculations using different \chiEFT{} nuclear
interaction models~\cite{Gazda:2016qva,Nogga:2019bwo}. Since the
\(\Lambda\) separation energies in \hnuc{4}{He} and \hnuc{4}{H} are
strongly correlated for each of the \nnlosim{} interactions, the model
uncertainty, quantified by the variance of their differences, is very
small, \(\sigma_{\mathrm{model}} \approx \unit[0.002(0.003)]{MeV}\)
for the \(0^+\)(\(1^+\)) state.

The uncertainty from the extrapolation procedure, indicated by the
error bars in Fig.~\ref{fig:CSB}, is larger than the model
uncertainty. This is particularly true for the $1^+$ states for which
our computations are limited to $\nm \leq 16$. Based on a correlation
study of \hnuc{4}{H,He} energies at different \nm{} we have estimated
a correlation coefficient $r=0.9$ between the \hnuc{4}{H,He}
extrapolation errors. This correlation implies that the corresponding
error in the difference of the two extrapolated energies becomes a
factor $\sqrt{10}$ smaller than with zero correlation (where the total
error would be the square root of the quadratic sum). We note that an
assignment of uncorrelated errors would in fact have been a much
stronger assumption. The resulting method uncertainties are on the
order of \(\sigma_{\mathrm{method}} \approx \unit[0.005(0.05)]{MeV}\)
for the \(0^+\)(\(1^+\)) state. In the future, it will be important to
perform similar uncertainty quantification studies with \YN{}
interaction models incorporating \CSB{} terms.

\section{Summary and outlook}
\label{sec:conclusion}
In this work we have used the \emph{ab initio} \YNCSM{} method to
study light hypernuclei up to \hnuc{5}{He} using realistic chiral
interactions as the only simulation input. In particular, we have made
a significant effort to quantify relevant \emph{nuclear} uncertainties
which we define as those that are related to the truncation and
calibration of the nuclear interaction model plus the error that can
be associated with the finite precision of the many-body solver.
Quantitative knowledge of these uncertainties is critical for future
research efforts in which hypernuclear structure data is used to
constrain the elusive \YN{} interaction.

The main findings and conclusions of this study are:
\begin{itemize}
\item{\textbf{A comprehensive study of nuclear interaction model
      uncertainties in hypernuclear observables.} We used the
    \nnlosim{} family of 42 realistic nuclear
    Hamiltonians~\cite{Carlsson:2015vda} to study the sensitivity of
    hypernuclear binding energies to the calibration and
    regularization of the nuclear interaction. We found that the
    \emph{model uncertainty} in the relevant $\Lambda$ separation
    energies ranges from \unit[20(100)]{keV} in \hnuc{3}{H}
    (\hnuc{4}{H,He}) to a few hundred keV in \hnuc{5}{He}.}
\item{\textbf{Significance of theoretical uncertainty quantification
      for constraining $\boldsymbol{\YN}$ interaction models.} We
    argued that the finite theoretical precision can be quantified and
    need to be taken into account in future efforts where spectra of
    light hypernuclei are used to constrain \YN{} interaction models.}
\item{\textbf{The \IR{} length scale of the truncated \YNCSM{} basis}
    was here established using both analytical and numerical
    arguments. This allowed us to apply rigorous \IR{} corrections to
    extract model-space converged results.}
\item{\textbf{Development of a Bayesian \IR{} extrapolation
      framework.} We have developed and used a fully Bayesian
    framework to perform the \IR{} extrapolation of \YNCSM{} results.
    This allowed the inclusion of both \LO{} and \NLO{} \IR{}
    corrections in the analysis, introducing various nuisance
    parameters with prior expectations but conditional on computed
    data. The method was validated for \hnuc{3}{H} and used to obtain
    converged results for light hypernuclei using the full family of
    \nnlosim{} interactions. This (extrapolation) \emph{method
      uncertainty} ranges from \unit[1(10)]{keV} for the ground states
    of \hnuc{3}{H} (\hnuc{4}{H,He}), to \unit[100]{keV} for the
    excited $1^+$ states in \hnuc{4}{H,He} and \unit[200--300]{keV} in
    \hnuc{5}{He} with the main limitation in precision originating in
    the computation restriction in \nm{}.}
\item{\textbf{The handling of correlated \IR{} errors.} In particular,
    we presented and applied a simple, stochastic model for the \IR{}
    corrections that allows to capture correlations between results
    obtained at different \IR{} length scales. This approach was
    critical in order to not overestimate the extrapolation errors.
    Furthermore, it can be straightforwardly applied to other \IR{}
    extrapolation studies.}
\item{\textbf{Theoretical precision of \CSB{} energy level splittings
      in $\boldsymbol{\hnuc{4}{H,He}}$.} We verified that the
    \(\Lambda\) separation energies in \hnuc{4}{H,He} are strongly
    correlated and that these correlations considerably reduce the
    theoretical uncertainties of the \CSB{} energy level splittings.
    In particular, we showed that the model uncertainty is very small,
    \(\sigma_{\mathrm{model}} \approx \unit[0.002(0.003)]{MeV}\) for
    the \(0^+\)(\(1^+\)) state. This precisely measured observable is
    therefore sensitive to properties of the \YN{} interaction, such
    as its poorly known charge dependence.}
\item{\textbf{Excessive \(\boldsymbol{\Lambda}\) separation energy in
      $\boldsymbol{\hnuc{5}{He}}$.} We have confirmed that the
    \loyn{}(\LamYN = \unit[600]{MeV}) yields too large \(\Lambda\)
    separation energy in \hnuc{5}{He}. However, we also found a large
    sensitivity of this observable to the \LamYN{} cutoff. Larger
    values of \(\LamYN\) seem to give a better agreement with the
    experimental value. Taking into account the considerable
    theoretical method and model uncertainties, we do not find a
    strong signal of deficiencies in the \loyn{} interaction.}
\end{itemize}

\begin{acknowledgments}
  We are grateful to Petr Navr\'{a}til for helpful advice on extending
  the nuclear \NCSM{} codes to hypernuclei, to Johann Haidenbauer, and
  Andreas Nogga for providing us with the input \LO{} Bonn--J\"{u}lich
  $\YN$ potentials used in the present work. The work of D.G.\ was
  supported by the Czech Science Foundation GA\v{C}R grants 19-19640S
  and 22-14497S, and by the Knut and Alice Wallenberg Foundation (PI:
  Jan Conrad). The research of T.Y.\ was performed at Chalmers through
  a PhD student partnership between the Swedish International
  Development Cooperation Agency (Sida), the Thailand International
  Development Cooperation Agency (TICA, and the Thailand Research Fund
  (TRF) coordinated by the International Science Program (ISP) at
  Uppsala University. The work of C.F.\ was supported by the Swedish
  Research Council (dnr.~2017-04234 and 2021-04507). Some of the
  computations and data handling were performed on resources provided
  by the Swedish National Infrastructure for Computing (SNIC) at C3SE
  (Chalmers) and NSC (Link\"oping) partially funded by the Swedish
  Research Council through grant agreement no. 2018-05973. Additional
  computational resources were supplied by the project
  ``e-Infrastruktura CZ'' (e-INFRA CZ LM2018140) supported by the
  Ministry of Education, Youth and Sports of the Czech Republic and
  IT4Innovations at Czech National Supercomputing Center under project
  number~OPEN-24-21~1892.
\end{acknowledgments}

\appendix*
\section{Bayesian inference parameters}
\label{app:hyperpars}
See Table~\ref{tab:hyperpars} for prior and posterior distributions
for the parameters used in the Bayesian infrared extrapolations of the
ground-state energies of \hnuc{3,4}{H} and \hnuc{4,5}{He} and the
$1^+$ excited states of \hnuc{4}{H,He}.
\begin{table*}[tb]
  \caption{Prior and posterior distributions for the parameters used
    in the Bayesian infrared extrapolations of the ground state
    energies of \hnuc{3,4}{H} and \hnuc{4,5}{He} and the $1^+$ excited
    states of \hnuc{4}{H,He}. The prior bounds for all parameters
    (except \epsbar{}) are given in the rows labeled ``prior''. For
    \epsbar{} we use a weakly informative inverse gamma distribution
    for which we present the 95\% credible interval. The posteriors
    are summarized by the median value of the distribution plus the
    68\% and 95\% credible intervals. The theoretical separation
    momenta, obtained from the medians of the binding energies, are
    shown in the final column for comparison with \kinf{} (see
    Section.~\ref{sec:irextrapol}).}
  \label{tab:hyperpars}
  \begin{ruledtabular}
    \begin{tabular}{c c c c c c c c c}
      \multicolumn{1}{c}{system}
      & &
      & \multicolumn{1}{c}{\Einf}
      & \multicolumn{1}{c}{\dEinf}
      & \multicolumn{1}{c}{\epsbar}
      & \multicolumn{1}{c}{$\rho$}
      & \multicolumn{1}{c}{\kinf}
      & \multicolumn{1}{c}{\ksepth}
      \\
      \hline
      \multirow{4}{*}{\hnuc{3}{H}} & prior &  & [-2.44,-2.29] & [0.00,0.05] & [0.21,9.27] & [0.01,0.99] & [0.03,0.26] \\
      & \multirow{3}{*}{posterior} & median  &  -2.391 &   0.007 &    0.28 &    0.98 &    0.13 &    0.09 \\
      & & 68\% CI  & [-2.391, -2.390] & [0.006, 0.008] & [ 0.20,  0.36] & [ 0.97,  0.99] & [ 0.12,  0.13] \\
      & & 95\% CI  & [-2.392, -2.389] & [0.005, 0.008] & [ 0.15,  0.48] & [ 0.94,  0.99] & [ 0.12,  0.14] \\
      \hline
      \multirow{4}{*}{\hnuc{4}{H}} & prior &  & [-11.48,-11.08] & [0.00,0.30] & [0.21,9.27] & [0.01,0.99] & [0.12,1.08] \\
      & \multirow{3}{*}{posterior} & median  &  -11.26 &    0.08 &    0.31 &    0.93 &    0.48 &    0.37 \\
      & & 68\% CI  & [-11.27, -11.25] & [ 0.07,  0.09] & [ 0.19,  0.52] & [ 0.71,  0.98] & [ 0.46,  0.51] \\
      & & 95\% CI  & [-11.29, -11.24] & [ 0.06,  0.11] & [ 0.13,  0.94] & [ 0.24,  0.99] & [ 0.42,  0.54] \\
      \hline
      \multirow{4}{*}{\hnuc{4}{He}} & prior &  & [-10.72,-10.32] & [0.00,0.30] & [0.21,9.27] & [0.01,0.99] & [0.12,1.08] \\
      & \multirow{3}{*}{posterior} & median  &  -10.48 &    0.08 &    0.30 &    0.93 &    0.48 &    0.36 \\
      & & 68\% CI  & [-10.49, -10.47] & [ 0.07,  0.10] & [ 0.19,  0.51] & [ 0.72,  0.98] & [ 0.46,  0.50] \\
      & & 95\% CI  & [-10.51, -10.46] & [ 0.06,  0.12] & [ 0.13,  0.93] & [ 0.25,  0.99] & [ 0.42,  0.53] \\
      \hline
      \multirow{4}{*}{\hnuc{5}{He}} & prior &  & [-35.48,-33.38] & [0.00,2.00] & [0.21,9.27] & [0.01,0.99] & [0.17,1.53] \\
      & \multirow{3}{*}{posterior} & median  &  -34.12 &    0.93 &    0.50 &    0.63 &    0.81 &    0.54 \\
      & & 68\% CI  & [-34.40, -33.95] & [ 0.75,  1.23] & [ 0.29,  0.96] & [ 0.25,  0.89] & [ 0.73,  0.88] \\
      & & 95\% CI  & [-34.95, -33.73] & [ 0.54,  1.74] & [ 0.18,  2.08] & [ 0.05,  0.97] & [ 0.62,  0.98] \\
      \hline
      \multirow{4}{*}{\hnuc{4}{H};$1^+$} & prior &  & [-10.62,-9.77] & [0.00,0.75] & [0.21,9.27] & [0.01,0.99] & [0.09,0.78] \\
      & \multirow{3}{*}{posterior} & median  &  -10.23 &    0.36 &    0.40 &    0.79 &    0.48 &    0.29 \\
      & & 68\% CI  & [-10.34, -10.14] & [ 0.26,  0.48] & [ 0.23,  0.77] & [ 0.36,  0.96] & [ 0.41,  0.55] \\
      & & 95\% CI  & [-10.53, -10.04] & [ 0.18,  0.66] & [ 0.15,  1.62] & [ 0.07,  0.99] & [ 0.32,  0.66] \\
      \hline
      \multirow{4}{*}{\hnuc{4}{He};$1^+$} & prior &  & [-9.97,-9.12] & [0.00,0.75] & [0.21,9.27] & [0.01,0.99] & [0.09,0.81] \\
      & \multirow{3}{*}{posterior} & median  &   -9.43 &    0.37 &    0.40 &    0.78 &    0.47 &    0.29 \\
      & & 68\% CI  & [-9.55, -9.33] & [ 0.28,  0.51] & [ 0.23,  0.76] & [ 0.36,  0.96] & [ 0.40,  0.55] \\
      & & 95\% CI  & [-9.76, -9.23] & [ 0.18,  0.68] & [ 0.15,  1.62] & [ 0.07,  0.99] & [ 0.31,  0.66] \\
    \end{tabular}
  \end{ruledtabular}
\end{table*}

\bibliography{refs}

\end{document}